\documentclass[10pt,a4paper]{article}
\usepackage{authblk}
\usepackage{amsmath}
\usepackage{amssymb}
\usepackage{amsthm}
\DeclareMathOperator{\arctanh}{arctanh}
\usepackage{esdiff}
\usepackage{color}
\usepackage{graphicx}
\usepackage{tikz}
\usetikzlibrary{arrows}
\usetikzlibrary{decorations}
\usepackage[numbers]{natbib}

\usepackage{stmaryrd}
\newcommand{\ljump}{\llbracket}
\newcommand{\rjump}{\rrbracket}
\newcommand{\jump}[1]{\ljump #1 \rjump}
\newcommand{\av}[1]{\langle #1 \rangle}
\newcommand{\bR}{\mathbf{R}}

\usepackage{epstopdf}

\title{Interfacial Cracks in Piezoelectric Bimaterials: \\ an approach based on Weight Functions and \\ Boundary Integral Equations}
\author[1]{L. Pryce}
\author[2]{L. Morini}
\author[1]{D. Andreeva}
\author[1]{A. Zagnetko}
\affil[1]{Department of Mathematics, Aberystwyth University,

Aberystwyth, Ceredigion, SY23 3BZ, Wales, UK.}
\affil[2]{IMT School for Advanced Studies, Piazza S.Francesco 19, 55100 Lucca, Italy.}
\date{\vspace{-5ex}}

\begin{document}
\maketitle
\begin{abstract}
The focus of this paper is on the analysis of a semi-infinite crack lying along a perfect interface in a piezoelectric bimaterial with arbitrary loading on the crack faces. 
Making use of the extended Stroh formalism for piezoelectric materials combined with Riemann-Hilbert formulation, general expressions are 
obtained for both symmetric and skew-symmetric weight functions associate with plane crack problems at the interface between 
dissimilar anisotropic piezoelectric media. The effect of the coupled electrical fields is incorporated in the derived original expressions
for the weight function matrices. These matrices are used together with Betti's reciprocity identity in order to obtain singular integral equations relating
the extended displacement and traction fields to the loading acting on the crack faces. In order to study the variation of the piezoelectric effect,
two different poling directions are considered. Examples are shown for both poling
directions with a number of mechanical and electrical loadings applied to the crack faces.\\

\end{abstract}

\section{Introduction}
Fracture in piezoelectric materials is an area of great interest due to their many applications in industry, for example, 
in piezoelectric insulators and actuators \citep{Lothe, Geis}. The problem of a static semi-infinite interfacial crack between 
dissimilar anisotropic piezoelectric materials under symmetric loading conditions has been studied in \citet{Suopiezo} using an approach 
based on the Stroh formalism \citep{Stroh} and Riemann-Hilbert formulation. As an alternative to this method, singular integral formulations for two-dimensional 
interfacial crack problems in piezoelectric bimaterials have been derived by means of approaches based on Green's function method \citep{GaoWang}. 
Applying this procedure, the displacements and the stresses are defined by integral relations involving the Green's
functions, for which explicit expressions are required. Although Green's functions for several crack problems in piezoelectric bimaterials 
have been derived \citep{Pan1, PanYuan}, their utilisation in evaluating physical displacements and stress fields on the
crack faces requires challenging numerical estimation of integrals for which convergence should be asserted
carefully. Moreover, both the complex variable formulation proposed by \citet{Suopiezo} and the approaches 
based on Green's function method work when the tractions applied on the discontinuity
surface are symmetric, but not in the case of asymmetric loading acting on the crack faces.

In this paper, we illustrate a general procedure for studying plane interfacial crack problems in anisotropic piezoelectric bimaterials
in the presence of a general non-symmetric mechanical and electrostatic loading distribution acting on the crack faces. 
The approach developed in \citet{Piccolroaz09} and \citet{Piccolroaz13}, based on weight function theory and Betti's reciprocity theorem, is extended to the case of
an interfacial crack in a piezoelectric bimaterial containing a perfect interface. Applying this method, the crack problem is also formulated in terms of
singular integral equations avoiding the use of Green's functions and the resulting challenging computations. 

In fracture mechanics the notion of weight function, defined as the stress intensity factor associated to a point load acting on the crack faces, was
originally introduced by \citet{Bueckner1970, Bueckner1, Bueckner2} and \citet{RiceWF}. This weight functions concept have been extended for studying 
crack problems in piezoelectric materials by \citet{McMeeking} and \citet{Ma}. An alternative formulation where the weight functions are defined
as the singular displacement field of the homogeneous traction free problem was proposed by \citet{WillisMovchan}. Recently, this weight functions definition
has been used in the derivation of stress intensity factors for both static and dynamic crack problems in isotropic and anisotropic bimaterials \citep{Piccolroaz09, Morini, Pryce1}
and in thermodiffusive elastic media \citep{Morini3}.
These weight functions have also been used in the derivation of singular integral equations relating 
interfacial tractions and crack displacement to the applied loadings on the crack faces \citep{Piccolroaz13,Morini2,AAG4}. In this paper, this method 
is generalised in order to study interfacial cracks in piezoelectric bimaterials. Explicit weight functions are derived for the class of transversely isotropic
piezoelectric bimaterials considering two different poling directions \citep{Hwupiezo}. Then these weight function matrices are used
together with the generalised Betti's identity to derive singular integral equations relating the tractions and displacements 
to the electric fields introduced by the piezoelectric effect. Finally, some simple illustrative examples of the application of the obtained singular integral
identities are reported.

Section 2 of the paper introduces the problem geometry and a number of preliminary results required for further analysis.
In Section 3 we derive weight functions for 
transversely isotropic piezoelectric bimaterials considering two different poling directions using the definition introduced by \citet{WillisMovchan}.  The problem is reduced 
to only contain the fields affected by the piezoelectric effect. In Section 4 the obtained weight function matrices are used together with Betti's identity in order to derive singular integral equations relating
the physical fields to the applied mechanical and electrical loadings on the crack faces. In Section 5 we present a number of examples 
for a variety of mechanical and electrical loadings. For one of the considered examples we also present the results from finite element
computations and compare their accuracy to those obtained from our singular integral equations.

\section{Problem formulation and preliminary results}
In this section we introduce the mathematical model used for the remainder of the paper. Some preliminary results concerning two-dimensional interfacial cracks in 
anisotropic piezoelectric bimaterials are reported.

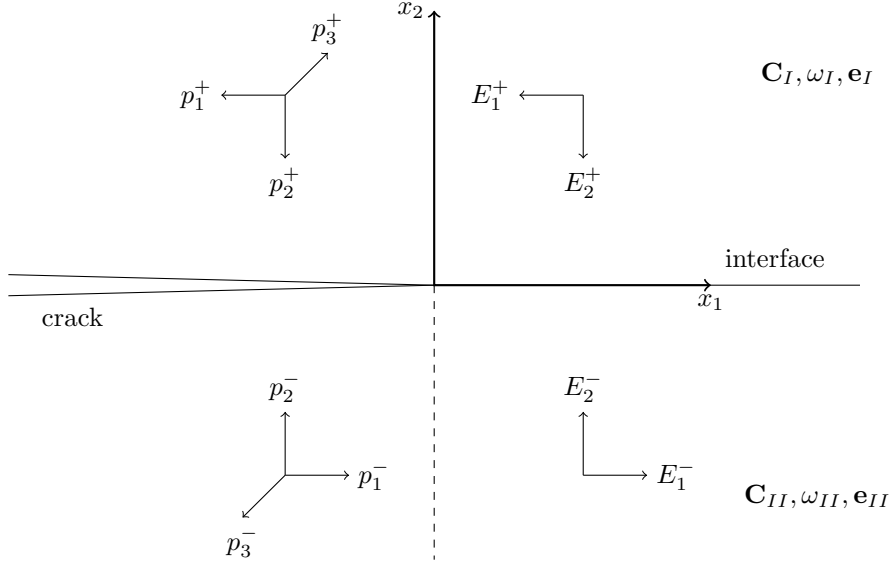
\begin{figure}
\begin{center}
\begin{tikzpicture}[scale=2.8]
\draw [-] (0,0) -- (2,0);
\draw [dashed] (0,0) -- (0,-1.3);
\draw [-] (-2,0.05) -- (0,0);
\draw [-] (-2,-0.05) -- (0,0);
\draw [thick, <->] (0, 1.3) -- (0,0) -- (1.3, 0);
\node [left] at (0,1.3) {$x_2$};
\node [below] at (1.3,0) {$x_1$};
\node [above] at (1.8,0.9) {$\mathbf{C}_I,\mathbf{\omega}_I,\mathbf{e}_I$};
\node [below] at (1.8,-0.9) {$\mathbf{C}_{II},\mathbf{\omega}_{II},\mathbf{e}_{II}$};
\node [below] at (-1.7,-0.06) {crack};
\node [above] at (1.6,0.04) {interface};
%
\draw[->] (0.7,0.9) -- (0.7,0.6);
\node[below] at (0.7,0.6) {$E_2^+$};
\draw[->] (0.7,0.9) -- (0.4,0.9);
\node[left] at (0.4,0.9) {$E_1^+$};
%
\draw[->] (-0.7,0.9) -- (-0.7,0.6);
\node[below] at (-0.7,0.6) {$p_2^+$};
\draw[->] (-0.7,0.9) -- (-1,0.9);
\node[left] at (-1,0.9) {$p_1^+$};
\draw[->] (-0.7,0.9) -- (-0.5,1.1);
\node[above] at (-0.5,1.1) {$p_3^+$};
%
\draw[->] (0.7,-0.9) -- (0.7,-0.6);
\node[above] at (0.7,-0.6) {$E_2^-$};
\draw[->] (0.7,-0.9) -- (1,-0.9);
\node[right] at (1,-0.9) {$E_1^-$};
%
\draw[->] (-0.7,-0.9) -- (-0.7,-0.6);
\node[above] at (-0.7,-0.6) {$p_2^-$};
\draw[->] (-0.7,-0.9) -- (-0.4,-0.9);
\node[right] at (-0.4,-0.9) {$p_1^-$};
\draw[->] (-0.7,-0.9) -- (-0.9,-1.1);
\node[below] at (-0.9,-1.1) {$p_3^-$};
\end{tikzpicture}
\caption{A semi-infinite crack along an interface between two dissimilar piezoelectric materials subject to the state of generalised plane strain and short circuit 
$(\varepsilon_{3}^{\pm}=E_{3}^{\pm}=0)$}
\label{geometry}
\end{center}
\end{figure}

We consider a semi-infinite crack lying along a perfect interface between two dissimilar piezoelectric half-planes, referred to as materials I and II. The crack occupies the 
region $\{x_1<0,x_2=0\}$, as illustrated in Figure \ref{geometry}. The perfect interface conditions in a piezoelectric bimaterial are continuity of displacements, traction, electric
potential and the electric charge. The loading along the crack faces, for $x_1<0$, is known and given by the functions 
\begin{equation}
p_j^\pm (x_1)=\sigma_{2j}(x_1,0^{\pm}),\quad\text{for }j=1,2,3,\quad p_4^\pm (x_1)=D_2(x_1,0^\pm),
\end{equation}
where $\sigma_{ij}$ and ${D_i}$ represent tractions and electrical displacements respectively.

Expressions for the stress fields and displacements for a plane semi-infinite interfacial crack between dissimilar piezoelectric media have been derived by \citet{Suopiezo}, 
using an approach based on Stroh formalism \citep{Stroh} and Riemann-Hilbert formulation. This method, which is applied in the paper in order to derive explicit weight functions for
piezoelectric bimaterials, is summarised in Section \ref{Strohform}.

In Section \ref{WFpiezo}, the definition of weight functions proposed in \citet{Piccolroaz07} and \citet{Morini} is extended to the case of interfacial cracks between dissimilar 
piezoelectric materials. Finally, in Section \ref{BETTIpiezo}, the expression for the Betti's formula generalized to piezoelectric media by \citet{Hadpiezo} is reported. Further in the text,
weight functions and Betti's formula are used for formulating the crack problem shown in Fig. \ref{geometry} in terms of singular integral equations.

\subsection{Stroh formalism for piezoelectric materials}
\label{Strohform}
Under the conditions of static deformations, the governing equations for a linear and generally anisotropic piezoelectric body are \citep{DunnTaya, Pan}:
\begin{equation}\label{stressinduction}
\sigma_{ij,i}=0,\qquad D_{i,i}=0,
\end{equation} 
where body forces are assumed to be zero. The strain $\epsilon$ and electric field $\mathbf{E}$, are defined by the gradients
\begin{equation}\label{strainfield}
\varepsilon_{ij}=\frac{1}{2}(u_{i,j}+u_{j,i}),\qquad E_i=-\phi_{,i},
\end{equation}
where $u_{i}$ are the components of the displacement field and $\phi$ is the electric potential.

The constitutive relations for anisotropic piezoelectric materials are given by \citet{WangZhou}  
\begin{equation}\label{piezorel}
\sigma_{ij}=C_{ijrs}\varepsilon_{rs}-e_{sji}E_s,\qquad D_i=\omega_{is}E_s+e_{irs}\varepsilon_{rs},
\end{equation}
where the tensors $C$, $\omega$ and $e$ represent the stiffness, permittivity and piezoelectricity respectively. Combining equations (\ref{stressinduction}), (\ref{strainfield}) 
and (\ref{piezorel}) the following relationships are obtained
\begin{equation}\label{piezogov}
(C_{ijrs}u_r+e_{sji}\phi)_{,si}=0,\qquad (-\omega_{is}\phi+e_{irs}u_r)_{,si}=0.
\end{equation}

Considering the semi-infinite interfacial crack problem shown in Fig. \ref{geometry}, and assuming
for both the upper and lower piezoelectric half-planes the state of generalised plane strain and short circuit $(\varepsilon_{3}=0$ and $E_{3}=0)$, 
a solution for equations (\ref{piezogov}) is derived by applying the procedure described by \citet{Suopiezo}. The solution
is sought in the form $(u_r,\phi)^T=\mathbf{a}f(z)$ where $z=x_1+px_2$. 
This yields the following eigenvalue problem
\begin{equation}\label{eigenpiezo}
[\mathbf{Q}+p(\mathbf{R}+\mathbf{R}^T)+p^2\mathbf{T}]\mathbf{A}=0
\end{equation}
For general, anisotropic piezoelectric media the matrices have the following form:
\[
\mathbf{Q}=\begin{pmatrix} C_{11}&C_{16}&C_{15}&e_{11}\\
C_{16}&C_{66}&C_{56}&e_{16}\\C_{15}&C_{56}&C_{55}&e_{15}\\e_{11}&e_{16}&e_{15}&-\omega_{11}\end{pmatrix},
\quad
\mathbf{R}=\begin{pmatrix} C_{16}&C_{12}&C_{14}&e_{16}\\
C_{66}&C_{26}&C_{46}&e_{12}\\C_{56}&C_{25}&C_{45}&e_{14}\\e_{21}&e_{26}&e_{25}&-\omega_{12}\end{pmatrix},
\quad
\mathbf{T}=\begin{pmatrix} C_{66}&C_{26}&C_{46}&e_{26}\\
C_{26}&C_{22}&C_{24}&e_{22}\\C_{46}&C_{24}&C_{44}&e_{24}\\e_{26}&e_{22}&e_{24}&-\omega_{22}\end{pmatrix}.
\]
The eight eigenvalues of \eqref{eigenpiezo} are found to be four pairs of complex conjugate

For the remainder of this paper $\mathbf{u}$ shall be used to the extended vector containing both the physical displacement and electric
potential given by $\mathbf{u}=(u_1,u_2,u_3,\phi)^T$. The extended traction vector, $\mathbf{t}$, is also introduced, given by $\mathbf{t}=(\sigma_{2i},D_2)^T$. It was 
shown in \citet{Suopiezo} that the extended displacements and traction are given by
\begin{equation}\label{ut}
\mathbf{u}=2\mathrm{Re}\sum_{\mu=1}^4 \mathbf{A}_\mu f_\mu(z_\mu),
\qquad
\mathbf{t}=2\mathrm{Re}\sum_{\mu=1}^4 \mathbf{L}_\mu f'_\mu(z_\mu).
\end{equation}
where
\[
L_{j\mu}=\sum_{r=1}^3[(C_{2jr1}+p_\mu C_{2jr2})A_{r\mu}] +(e_{1j2}+p_\mu e_{2j2})A_{4\mu},\quad\text{for }j=1,2,3,
\]\[
L_{4\mu}=\sum_{r=1}^3[(e_{2r1}+p_\mu e_{2r2})A_{r\mu}]-(\omega_{12}+p_\mu \omega_{22})A_{4\mu}.
\]
The values of $z_\mu$ are given by $z_\mu=x_1+p_\mu x_2$, where the eigenvalues, $p_\mu$, are taken to be those with positive imaginary part and the
matrix $\mathbf{A}_\mu$ are the corresponding eigenvectors. The surface admittance tensor $\mathbf{B}=i\mathbf{AL}^{-1}$ is introduced and for further analysis 
the following bimaterial matrices are defined
\begin{equation}\label{HW}
\mathbf{H}=\mathbf{B}_I+\bar{\mathbf{B}_{II}},\qquad\mathbf{W}=\mathbf{B}_I-\bar{\mathbf{B}_{II}}.
\end{equation} 

Assuming vanishing tractions and electrical displacement along the crack faces for $x_1<0$, and using expressions (\ref{ut}), in \citet{Suopiezo} the 
following Riemann-Hilbert problem was found to exist along the negative $x_1$- axis
\begin{equation}\label{RH}
\mathbf{h}^+(x_1)+\bar{\mathbf{H}}^{-1}\mathbf{H}\mathbf{h}^-(x_1)=0,\quad -\infty<x_1<0.
\end{equation}
A solution is found in the form $\mathbf{h}(z)=\mathbf{w}z^{-\frac{1}{2}+i\epsilon}$ where the branch cut is situated along the negative real axis (which is the crack line). Inserting 
this solution into equation \eqref{RH} yields the eigenvalue problem
\begin{equation}\label{Heigen}
\bar{\mathbf{H}}\mathbf{w}=e^{2\pi\epsilon}\mathbf{H}\mathbf{w}.
\end{equation}
Four pairs of eigenvectors and eigenvalues are found from (\ref{Heigen}). They are
\begin{equation}
(\epsilon,\mathbf{w}),\: (-\epsilon,\bar{\mathbf{w}}),\: (-i\kappa,\mathbf{w}_3),\: (i\kappa,\mathbf{w}_4).
\end{equation}

Imposing the continuity of tractions and electrical displacement along the interface ahead of the crack tip,
the following formula for determining the extended traction vector $\mathbf{t}$ is derived \citep{Suopiezo}:
\begin{equation}\label{RHnoHom}
\mathbf{h}^+(x_1)+\bar{\mathbf{H}}^{-1}\mathbf{H}\mathbf{h}^-(x_1)=\mathbf{t}(x_1),\quad 0<x_1<\infty.
\end{equation}
The extended traction vector, derived by means of equation (\ref{RHnoHom}), is therefore given by
\begin{equation}\label{traction}
\mathbf{t}(x_1)=(2\pi x_1)^{-\frac{1}{2}}[Kx_1^{i\epsilon}\mathbf{w}+\bar{K}x_1^{-i\epsilon}\bar{\mathbf{w}}+ K_3x_1^\kappa\mathbf{w}_3+K_4x_1^{-\kappa}\mathbf{w}_4],
\end{equation}
where $K=K_1+iK_2$. The stress intensity factor vector is given by $\mathbf{K}=(K,\bar{K},K_3,K_4)^T$.

The extended displacement jump across the crack, given by $\jump{\mathbf{u}}(x_1)=\mathbf{u}^+(x_1)-\mathbf{u}^-(x_1)$, was found to be
\begin{align}\label{displacementjump}
\jump{\mathbf{u}}(x_1)=(\mathbf{H}+\bar{\mathbf{H}})\sqrt{\frac{(-x_1)}{2\pi}}\bigg{[}\frac{K(-x_1)^{i\epsilon}\mathbf{w}}{(1+2i\epsilon)\cosh\pi\epsilon}&+\frac{\bar{K}(-x_1)^{-i\epsilon}\bar{\mathbf{w}}}{(1-2i\epsilon)\cosh\pi\epsilon}\nonumber\\
&+\frac{K_3(-x_1)^{\kappa}\mathbf{w}_3}{(1+2\kappa)\cos\pi\kappa}+\frac{K_4(-x_1)^{-\kappa}\mathbf{w}_4}{(1-2\kappa)\cos\pi\kappa}
\bigg{]}.
\end{align}
Using these expressions for the traction and displacement jump the following equation can be used to find the energy release rate at the crack tip
\begin{equation}
G=\frac{1}{2\nu}\int_0^\nu\mathbf{t}^T(\Delta-r)[\mathbf{u}](r)\mathrm{d}r,
\end{equation}
for an arbitrary value of $\nu$. Using expressions (\ref{traction}) and (\ref{displacementjump}) it can be shown that
\begin{equation}\label{ERR}
G=\frac{\bar{\mathbf{w}}^T(\mathbf{H}+\bar{\mathbf{H}})\mathbf{w}}{4\cosh^2\pi\epsilon}|K|^2
+\frac{\mathbf{w}_3^T(\mathbf{H}+\bar{\mathbf{H}})\mathbf{w}_4}{4\cos^2\pi\kappa}K_3 K_4.
\end{equation}

\subsection{Weight functions for piezoelectric bimaterials: definition}
\label{WFpiezo}
Weight functions, introduced in \citet{Bueckner1, Bueckner2}, are functions whose weighted integrals can be used in the derivation of important fracture
parameters, such as the stress intensity factors. For elastic materials \citet{WillisMovchan} introduced a weight function given by the singular displacement
field corresponding to a homogeneous, traction-free problem similar to Fig. \ref{geometry} with the crack occupying the region $x_1>0$ and 
the perfect interface lying along the region $x_1<0$. This concept of a weight function has been adopted
in the works of \citet{Piccolroaz07} and \citet{Morini} for studying interfacial cracks in isotropic and anisotropic bimaterials, 
respectively. Here, this approach is extended to piezoelectric materials.

The weight function for piezoelectric materials is given by the extended singular displacement field $\mathbf{U}$
incorporating both displacement and electric potential. The symmetric and skew-symmetric parts of the weight function across the plane
$x_{2}=0$ are given by 
\begin{equation}
\jump{\mathbf{U}}(x_1)=\mathbf{U}(x_1,0^+)-\mathbf{U}(x_1,0^-),
\end{equation}\begin{equation}
\av{\mathbf{U}}(x_1)=\frac{1}{2}\left[ \mathbf{U}(x_1,0^+)+\mathbf{U}(x_1,0^-)\right].
\end{equation}
To satisfy the perfect interface conditions it is clear that $\jump{\mathbf{U}}=0$ for $x_1<0$.

The extended traction field corresponding to the extended displacement $\mathbf{U}$, is denoted $\mathbf{\Sigma}$. The following 
Riemann-Hilbert problem is found along the positive portion of the $x_1$-axis
%
\begin{equation} \label{RHweight}
\mathbf{h}^+(x_1)+\bar{\mathbf{H}}^{-1}\mathbf{H}\mathbf{h}^-(x)=0,\quad 0<x_1<\infty.
\end{equation}
A solution for $\mathbf{h}(z)$ is now sought in the form $\mathbf{h}=\mathbf{v}^{-\frac{3}{2}+i\epsilon}$. The branch cut of $\mathbf{h}$ is situated along the positive part of the $x_1$axis.
Inserting this solution into \eqref{RHweight} yields the following eigenvalue problem
\begin{equation}\label{Hbareigen}
\bar{\mathbf{H}}\mathbf{v}=e^{-2\pi\epsilon}\mathbf{H}\mathbf{v}.
\end{equation}
It is immediately clear by comparing equation \eqref{Heigen} to \eqref{Hbareigen} that $\mathbf{v}=\bar{\mathbf{w}}$, $\mathbf{v}_3=\mathbf{w}_4$ and $\mathbf{v}_4=\mathbf{w}_3$.

Along the negative part of the real axis $\mathbf{\Sigma}$ is given by
\begin{equation}
\mathbf{h}^+(x_1)+\bar{\mathbf{H}}^{-1}\mathbf{H}\mathbf{h}^-(x)=\mathbf{\Sigma}(x_1),\quad -\infty<x_1<0.
\end{equation}
Therefore the extended traction vector corresponding to the weight function $\mathbf{U}$ is given by
\begin{equation}
\mathbf{\Sigma}(x_1)=\frac{(-x_1)^{-3/2}}{2\sqrt{2\pi}}\left[C(-x_1)^{i\epsilon}\bar{\mathbf{w}}+\bar{C}(-x_1)^{-i\epsilon}\mathbf{w} + C_3(-x_1)^{-\kappa}\mathbf{w}_3 + C_4(-x_1)^{\kappa}\mathbf{w}_4\right],
\end{equation}
where $C=C_1+iC_2$, $C_3$ and $C_4$ are constants defined in the same manner as the stress intensity factors for the physical problem.

Let us introduce the Fourier transform of a generic function $f$ with respect to the variable $x_{1}$, defined as follows:
\begin{equation}
 \hat{f}(\xi)=\mathcal{F}[f(x_{1})]=\int_{-\infty}^{+\infty}f(x_{1})e^{i\xi x_{1}}d x_{1}, \quad  f(x_{1})=\mathcal{F}^{-1}[\hat{f}(\xi)]=\frac{1}{2\pi}\int_{-\infty}^{+\infty}\hat{f}(\xi)e^{-i\xi x_{1}}d \xi.
\end{equation}
For anisotropic materials it was shown in \citet{Morini} that the Fourier transforms of the symmetric and skew-symmetric parts of the weight functions are related 
to $\hat{\Sigma}$ in the following manner
\begin{equation}\label{Usymft}
\jump{\hat{\mathbf{U}}}^+(\xi)=\frac{1}{|\xi|}(i\mathrm{sign}(\xi)\mathrm{Im}(\mathbf{H})- \mathrm{Re}(\mathbf{H}))\hat{\Sigma}^-(\xi),
\end{equation}
\begin{equation}\label{Uskewft}
\av{\hat{\mathbf{U}}}(\xi)=\frac{1}{2|\xi |}(i\mathrm{sign}(\xi)\mathrm{Im}(\mathbf{W})- \mathrm{Re}(\mathbf{W}))\hat{\Sigma}^-(\xi),
\end{equation}

As the method used in \citet{Morini} was for general matrices $\mathbf{H}$ and $\mathbf{W}$, it is immediately clear that these results also hold for piezoelectric materials.

\subsection{The generalised Betti formula}
\label{BETTIpiezo}
In this part of the paper we consider the Betti identity in the context of a semi-infinite crack in a piezoelectric bimaterial. Originally used to relate two different sets of
displacement and traction fields satisfying the equilibrium equations \citep{WillisMovchan, Piccolroaz07}, 
this approach was extended to study piezoelectric materials (with electric potential and electric displacement) by \citet{Hadpiezo}. 
The derivation of Betti formula for piezoelectric solids is briefly reported in Appendix \ref{piezoBetti}. This integral formula is now used to form a relationship between 
the physical fields and the weight function introduced in the previous part of the paper. 

Applying the generalised Betti formula derived in \citet{Hadpiezo} to the semi-circular domain occupying the upper half plane
in the model illustrated in Fig. \ref{geometry}, the following integral relation is found
\begin{equation}\label{bettiup}
\int_{x_2=0^+}\left[\mathbf{R}\mathbf{U}(x_1'-x_1,0^+)\cdot \mathbf{t}(x_1,0^+) - \mathbf{R}\mathbf{\Sigma}(x_1'-x_1,0^+)\cdot \mathbf{u}(x_1,0^+)\right]\mathrm{d}x_1=0,
\end{equation}
where $\mathbf{u}=(u_1,u_2,u_3,\phi)^T$ and $\mathbf{t}=(\sigma_{21},\sigma_{22}, \sigma_{23}, D_2)^T$ (see the Appendix for details) and $\mathbf{R}$ is given by
\[
\mathbf{R}=\begin{pmatrix}-1&0&0&0\\0&1&0&0\\0&0&-1&0\\0&0&0&-1\end{pmatrix}.
\]
The equivalent equation for a semi-circular domain in the lower half-plane gives
\begin{equation}\label{bettidown}
\int_{x_2=0^-}\left[\mathbf{R}\mathbf{U}(x_1'-x_1,0^-)\cdot \mathbf{t}(x_1,0^-) - \mathbf{R}\mathbf{\Sigma}(x_1'-x_1,0^-)\cdot \mathbf{u}(x_1,0^-)\right]\mathrm{d}x_1=0.
\end{equation}
Subtracting equation \eqref{bettidown} from \eqref{bettiup} yields the following relationship
\begin{equation}\label{betconv}
\mathbf{R}\jump{\mathbf{U}}\ast\mathbf{t}^{(+)}-\mathbf{R}\mathbf{\Sigma}^{(-)}\ast\jump{\mathbf{u}}=-\mathbf{R}\jump{\mathbf{U}}\ast\av{\mathbf{p}}-\mathbf{R}\av{\mathbf{U}}\ast\jump{\mathbf{p}},
\end{equation}
where $\ast$ represents the convolution with respect to $x_1$ and $^{(\pm)}$ is used to represent the restriction of a function to the positive or negative portion of the $x_1$-axis respectively. 
It can be easily deduced that in equation \eqref{betconv} the contribution to the generalised traction vector defined on the negative semi-axis $x_{1}<0$ is
given by the loading functions $\mathbf{t}^{(-)}=(\sigma_{21}^{(-)},\sigma_{22}^{(-)}, \sigma_{23}^{(-)}, D_2^{(-)})^T=(p_{1},p_{2},p_{3}, p_4)^T=\mathbf{p}$, and the 
symmetrical and skew-symmetrical part of the load, respectively $\av{\mathbf{p}}$ and $\jump{\mathbf{p}}$, are defined as follows \citep{Piccolroaz13}: 
\begin{equation}
 \av{\mathbf{p}}=\frac{1}{2}(\mathbf{p}^{+}+\mathbf{p}^{-}), \quad \jump{\mathbf{p}}=\mathbf{p}^{+}-\mathbf{p}^{-}.
\end{equation}

Applying the Fourier transform to \eqref{betconv} gives the following relationship
\begin{equation}\label{bettipiezo}
\jump{\hat{\mathbf{U}}}^T\mathbf{R}\hat{\mathbf{t}}^+ - (\hat{\mathbf{\Sigma}}^-)^T\mathbf{R}\jump{\hat{\mathbf{u}}} = -\jump{\hat{\mathbf{U}}}^T\mathbf{R}\av{\hat{\mathbf{p}}} -\av{\hat{\mathbf{U}}}^T\mathbf{R}\jump{\hat{\mathbf{p}}}.
\end{equation}

In the next Sections, explicit expressions for the weight function matrices \eqref{Usymft} and \eqref{Uskewft} are derived and used together with the the generalised Betti identity \eqref{bettipiezo}
for formulating the considered interface crack problem in terms of singular integral equations. Since the bimaterial matrices $\mathbf{H}$ and $\mathbf{W}$ involved in the 
weight functions \eqref{Usymft} and \eqref{Uskewft} depend on the surface admittance tensors of both piezoelectric half-planes (see definition \eqref{HW}), in order to derive explicit
expressions for these matrices the  solution of the Stroh's eigenvalue problem \eqref{eigenpiezo} is needed. In the general fully anisotropic case, this eigenvalue problem  
must be solved numerically. Nevertheless, exact algebraic expressions of Stroh's eigenvalues and eigenvectors have been obtained for the class of transversely isotropic piezoelectric 
materials in \cite{Suopiezo, Hwupiezo, OuWu} and \cite{Park}. This class of materials has practical significance, because many poled ceramics that are actually in use fall into this
category. The Stroh matrices and surface admittance tensors for transversely isotropic piezoelectric materials with poling direction parallel to $x_{2}$ and $x_{3}$ axes are
reported in the Appendix. Further in the text, these results will be used together with expressions \eqref{Usymft} and \eqref{Uskewft} for deriving explicit weight function matrices 
corresponding to interfacial cracks between dissimilar transversely isotropic piezoceramics.

\section{Weight functions}
\label{WFsec}
Piezoelectric materials occupying both lower and upper half-planes in Figure \ref{geometry} are assumed to be transversely isotropic. For
simplicity, poling direction is assumed to be parallel to the $x_{2}$ and $x_{3}$ axes, respectively.
Using eigenvalue matrices and surface admittance tensors in the forms reported in the Appendix, explicit weight functions are derived
for both these cases.

\subsection{Poling direction parallel to the $x_2-$axis}
Poling direction directed along the $x_{2}-$axis is assumed for both upper and lower piezoelectric half-planes. Considering the geometry 
of the model shown in Figure \ref{geometry}, it is easy to observe that in this case the poling direction is perpendicular to the crack plane.
Under these conditions the Mode III component of the solution decouple from Modes I and II and the piezoelectric effect \citep{Hwupiezo, OuWu, Park}, 
which further in the text we will refer to as Mode IV. 
This means that the antiplane tractions and displacement have no dependency on the electric field and therefore behave in the same way as they would
in an elastic material with no piezoelectric effect. The stiffness, permittivity and piezoelectric tensors corresponding to this case are reported in the 
Appendix together with explicit forms of the matrices involved in the decoupled part of the eigenvalue problem (\ref{eigenpiezo}).

The remainder of this section of the paper considers only the in-plane components and electrical effects.
That is: $\mathbf{u}=(u_1,u_2,\phi)^T$ and $\mathbf{t}=(\sigma_{21},\sigma_{22},D_2)^T$. The surface admittance tensor $\mathbf{B}$ then becomes
\begin{equation}
\mathbf{B}=\begin{pmatrix}
B_{11}&iB_{12}&iB_{14}\\
-iB_{12}&B_{22}&B_{24}\\
-iB_{14}&B_{24}&B_{44}
\end{pmatrix}.\end{equation}
The expressions for the components of the matrix $\mathbf{B}$, found by \citet{Hwupiezo} for the two-dimensional state of generalized plane
strain and open circuit, are quoted in the Appendix of the paper. With an expression for $\mathbf{B}$ 
it is now possible to construct the bimaterial matrices required. The bimaterial matrices $\mathbf{H}$ and $\mathbf{W}$ can be written as
\begin{equation}\label{HW3x3}
\mathbf{H}=\begin{pmatrix}
H_{11}&iH_{12}&iH_{14}\\
-iH_{12}&H_{22}&H_{24}\\
-iH_{14}&H_{24}&H_{44}
\end{pmatrix}, \quad
\mathbf{W}=\begin{pmatrix}
W_{11}&iW_{12}&iW_{14}\\
-iW_{12}&W_{22}&W_{24}\\
-iW_{14}&W_{24}&W_{44}
\end{pmatrix}.\end{equation}
Note that explicit expressions for the components of matrices \eqref{HW3x3} are also reported in the Appendix.

Knowing the structure of the bimaterial matrix $\mathbf{H}$ it is now possible to find expressions for the traction field, $\Sigma$ using
the eigenvalue problem \eqref{Heigen}. From \eqref{HW3x3}$_{(1)}$ it is only necessary to find three sets of eigenvalues and eigenvectors. They have the form
\[
(\epsilon,\mathbf{w}),\: (-\epsilon,\bar{\mathbf{w}}),\: (i\kappa,\mathbf{w}_4).
\]
As the Mode III components of the solutions have decoupled and behave purely elastically it is expected that $\kappa=0$.
Therefore it is expected that the eigenvalues are given by two non-zero real valued numbers, with the same magnitude but differing in sign, and 0.
With these particular eigenvalues and eigenvectors the expression for $\Sigma$ is given by
\begin{equation}\label{Sigma3x3}
\mathbf{\Sigma}(x_1)=\frac{(-x_1)^{-3/2}}{2\sqrt{2\pi}}\left[C(-x_1)^{i\epsilon}\bar{\mathbf{w}} + \bar{C}(-x_1)^{-i\epsilon}\mathbf{w} + C_4\mathbf{w}_4\right].
\end{equation}

To find the eigenvalues from \eqref{Heigen} the following equation must be solved
\begin{equation}
\label{Heigen3X3}
||\bar{\mathbf{H}}-e^{2\pi\epsilon}\mathbf{H}||=0.
\end{equation}
Substituting \eqref{HW3x3}$_{(1)}$ in \eqref{Heigen3X3}  the following equation is derived
\begin{equation}
(1-e^{2\pi\epsilon})[(1-e^{2\pi\epsilon})^2H_{11}(H_{22}H_{44}-H_{24}^2) - (1+e^{2\pi\epsilon})^2(H_{12}^2H_{44} + H_{14}^2H_{24} - 2H_{12}H_{14}H_{24})]=0.
\end{equation}
As expected solving the equation $1-e^{2\pi\epsilon}=0$ yields the eigenvalue 0. The other eigenvalues are given by 
\begin{equation}
\pm\epsilon=\pm\frac{1}{2\pi}\ln\left(\frac{1-\beta}{1+\beta}\right),
\end{equation}
where
\[\beta^2=\frac{B}{A},\quad A=H_{11}(H_{22}H_{44}-H_{24}^2),\quad B=2(H_{12}^2H_{44} + H_{14}^2H_{24} - 2H_{12}H_{14}H_{24}).\]

Using these eigenvalues it is possible to find expressions for the eigenvectors $\mathbf{w}$ and $\mathbf{w_4}$.
The expressions chosen here are made for notational convenience. The expression for $\mathbf{w_4}$ is
\begin{equation}
\mathbf{w_4}=\frac{1}{2}\begin{pmatrix}
0\\H_{14}\\-H_{12}
\end{pmatrix}.
\end{equation}
There are three possible expressions for the eigenvector $\mathbf{w}$. They are
\begin{equation}\label{threew}
\mathbf{w}=\frac{1}{2}\begin{pmatrix}
-i\beta(H_{24}^2-H_{22}H_{44})\\
H_{44}H_{12}-H_{14}H_{24}\\
H_{14}H_{22}-H_{24}H_{12}
\end{pmatrix},\text{ or   }
\frac{1}{2}\begin{pmatrix}
-i\beta(H_{14}H_{22}-H_{12}H_{24})\\
\beta^2H_{11}H_{24}-H_{12}H_{14}\\
H_{12}^2-\beta^2H_{11}H_{22}
\end{pmatrix},\text{ or   }
\frac{1}{2}\begin{pmatrix}
-i\beta(H_{14}H_{24}-H_{12}H_{44})\\
\beta^2H_{11}H_{44}-H_{14}^2\\
H_{12}H_{14}-\beta^2H_{11}H_{24}
\end{pmatrix}.
\end{equation}
For the remainder of this paper the first representation of $\mathbf{w}$ from equation \eqref{threew} shall be used.

Using \eqref{Sigma3x3} it is possible, using the method described in \citet{Piccolroaz09}, to construct three independent traction vectors using the following three cases: 
\begin{enumerate}
\item $C_1=1,C_2=C_4=0$,
\item $C_2=1,C_1=C_4=0$,
\item $C_4=1,C_1=C_2=0$.
\end{enumerate}
Using \eqref{threew} the three traction vectors obtained are 
\begin{equation}\label{sigma1}
\mathbf{\Sigma}^1(x_1)=\frac{(-x_1)^{-3/2}}{2\sqrt{2\pi}}\begin{pmatrix}
i\beta(H_{24}^2-H_{22}H_{44})[(-x_1)^{i\epsilon} - (-x_1)^{-i\epsilon}]\\
(H_{44}H_{12}-H_{14}H_{24})[(-x_1)^{i\epsilon} + (-x_1)^{-i\epsilon}]\\
(H_{14}H_{22}-H_{24}H_{12})[(-x_1)^{i\epsilon} + (-x_1)^{-i\epsilon}]
\end{pmatrix},
\end{equation}
\begin{equation}\label{sigma2}
\mathbf{\Sigma}^2(x_1)=\frac{(-x_1)^{-3/2}}{2\sqrt{2\pi}}\begin{pmatrix}
-\beta(H_{24}^2-H_{22}H_{44})[(-x_1)^{i\epsilon} + (-x_1)^{-i\epsilon}]\\
i(H_{44}H_{12}-H_{14}H_{24})[(-x_1)^{i\epsilon} - (-x_1)^{-i\epsilon}]\\
i(H_{14}H_{22}-H_{24}H_{12})[(-x_1)^{i\epsilon} - (-x_1)^{-i\epsilon}]
\end{pmatrix},
\end{equation}
\begin{equation}\label{sigma4}
\mathbf{\Sigma}^4(x_1)=\frac{(-x_1)^{-3/2}}{2\sqrt{2\pi}}\begin{pmatrix}
0\\H_{14}\\-H_{12}\end{pmatrix}.
\end{equation}
Here, a superscript 4 has been used instead of 3 in equation \eqref{sigma4} so as not to confuse this with the Mode III components which have already been decoupled. 

In order to calculate explicit expressions for $\jump{\hat{\mathbf{U}}}^+$ and $\av{\hat{\mathbf{U}}}$ it is necessary to find the Fourier transforms
of \eqref{sigma1}, \eqref{sigma2} and \eqref{sigma4}
\begin{equation}\label{sigmaft1}
\hat{\mathbf{\Sigma}}^1(\xi)=\frac{\sqrt{2}(i\xi)^{1/2}}{(1+4\epsilon^2)\sqrt{\pi}}
\begin{pmatrix}
i\beta(H_{24}^2-H_{22}H_{44})\left[ (-\frac{1}{2}-i\epsilon)\Gamma (\frac{1}{2}+i\epsilon)(i\xi)^{-i\epsilon} - (-\frac{1}{2}+i\epsilon)\Gamma (\frac{1}{2}-i\epsilon)(i\xi)^{i\epsilon}\right]\\
(H_{44}H_{12}-H_{14}H_{24})\left[ (-\frac{1}{2}-i\epsilon)\Gamma (\frac{1}{2}+i\epsilon)(i\xi)^{-i\epsilon} + (-\frac{1}{2}+i\epsilon)\Gamma (\frac{1}{2}-i\epsilon)(i\xi)^{i\epsilon}\right]\\
(H_{14}H_{22}-H_{24}H_{12})\left[ (-\frac{1}{2}-i\epsilon)\Gamma (\frac{1}{2}+i\epsilon)(i\xi)^{-i\epsilon} + (-\frac{1}{2}+i\epsilon)\Gamma (\frac{1}{2}-i\epsilon)(i\xi)^{i\epsilon}\right]
\end{pmatrix},
\end{equation}
\begin{equation}\label{sigmaft2}
\hat{\mathbf{\Sigma}}^2(\xi)=\frac{\sqrt{2}(i\xi)^{1/2}}{(1+4\epsilon^2)\sqrt{\pi}}
\begin{pmatrix}
-\beta(H_{24}^2-H_{22}H_{44})\left[ (-\frac{1}{2}-i\epsilon)\Gamma (\frac{1}{2}+i\epsilon)(i\xi)^{-i\epsilon} + (-\frac{1}{2}+i\epsilon)\Gamma (\frac{1}{2}-i\epsilon)(i\xi)^{i\epsilon}\right]\\
i(H_{44}H_{12}-H_{14}H_{24})\left[ (-\frac{1}{2}-i\epsilon)\Gamma (\frac{1}{2}+i\epsilon)(i\xi)^{-i\epsilon} - (-\frac{1}{2}+i\epsilon)\Gamma (\frac{1}{2}-i\epsilon)(i\xi)^{i\epsilon}\right]\\
i(H_{14}H_{22}-H_{24}H_{12})\left[ (-\frac{1}{2}-i\epsilon)\Gamma (\frac{1}{2}+i\epsilon)(i\xi)^{-i\epsilon} - (-\frac{1}{2}+i\epsilon)\Gamma (\frac{1}{2}-i\epsilon)(i\xi)^{i\epsilon}\right]
\end{pmatrix},
\end{equation}
\begin{equation}\label{sigmaft4}
\hat{\mathbf{\Sigma}}^4(\xi)=\frac{(i\xi)^{1/2}}{\sqrt{2}}
\begin{pmatrix}
0\\-H_{14}\\H_{12}
\end{pmatrix}.
\end{equation}
With these expressions it is now possible to use a 3x3 matrix whose columns are the three linearly independent traction vectors found along with equations \eqref{Usymft} and \eqref{Uskewft} to find expressions for $\jump{\mathbf{U}}$ and $\av{\mathbf{U}}$ \citep{Piccolroaz09}.

\subsection{Poling direction parallel to the $x_3-$axis}
Observing Figure \ref{geometry}, it can be noted that in the case where both the
upper and lower piezoelectric half-planes are assumed to be poled along the $x_{3}-$axis, the poling 
axis coincides with the crack front. For this particular case it is possible to decouple
the Mode I and Mode II components of the displacement and stress fields from the Mode III
fields and piezoelectric effects on the material \citep{ OuWu,Hwupiezo}. This means that the in-plane fields will behave similarly
to those for purely elastic materials with no piezoelectric behaviour. Also for this case, the explicit form for the stiffness, permittivity and
piezoelectric tensors are reported in the Appendix.

In the remainder of this section only the out-of-plane and piezoelectric components are considered. 
That is: $\mathbf{u}=(u_3,\phi)^T$ and $\mathbf{t}=(\sigma_{23},D_2)^T$. The surface admittance tensor then becomes
\begin{equation}
\mathbf{B}=\begin{pmatrix} B_{33} & B_{34} \\ B_{34} & B_{44} \end{pmatrix}.
\end{equation}
Consequently, the bimaterial matrices $\mathbf{H}$ and $\mathbf{W}$ can be computed and have the form
\begin{equation}\label{HW2X2}
\mathbf{H}=\begin{pmatrix}H_{33} & H_{34} \\ H_{34} & H_{44} \end{pmatrix},\qquad
\mathbf{W}=\begin{pmatrix} \delta_3 H_{33} & \gamma H_{34} \\ \gamma H_{34} & \delta_4 H_{44} \end{pmatrix},
\end{equation}
where:
\[H_{\alpha\beta}=[B_{\alpha\beta}]_I + [B_{\alpha\beta}]_{II},\quad\text{for }\alpha,\beta=3,4,\]
\[ \delta_\alpha = \frac{[B_{\alpha\alpha}]_I - [B_{\alpha\alpha}]_{II}}{[B_{\alpha\alpha}]_I + [B_{\alpha\alpha}]_{II}},\quad\text{for }\alpha=3,4,\]
\[ \gamma = \frac{[B_{34}]_I - [B_{34}]_{II}}{H_{34}}.\]
Explicit expressions for the components of $B_{\alpha\beta}$, are given in the Appendix of the paper.

In order to obtain the weight functions for the materials considered here the Riemann-Hilbert problem \eqref{RH} must again be considered.
For this case the bimaterial matrix $\mathbf{H}$ has no imaginary part, and then substituting expression \eqref{HW2X2}$_{(1)}$ into
equation \eqref{RH} we get
\begin{equation}\label{RHreal}
\mathbf{h}^+(x_1)+\mathbf{h}^-(x)=0,\quad -\infty<x_1<0.
\end{equation}
For this special case it was shown in \citet{Suopiezo} that the extended traction along the interface and displacement jump across the crack are given by
\begin{equation}
\mathbf{t}(x_1)=(2\pi x_1)^{-\frac{1}{2}}\begin{pmatrix}K_3\\K_4\end{pmatrix},\qquad\text{for }x_1>0,
\end{equation}
\begin{equation}
\jump{\mathbf{u}}(x_1)= \left( \frac{2(-x_1)}{\pi}\right)^{\frac{1}{2}}\mathbf{H}\begin{pmatrix}K_3\\K_4\end{pmatrix}\qquad \text{for }x_1<0.
\end{equation}

Knowing the traction fields makes it possible to evaluate the weight function $\mathbf{U}$ and its corresponding traction $\mathbf{\Sigma}$ for these 
particular materials. The expression for $\mathbf{\Sigma}$ is
\begin{equation}\label{parx3sigma}
\mathbf{\Sigma}(x_1)=\frac{(-x_1)^{-\frac{3}{2}}}{\sqrt{2\pi}}\begin{pmatrix}C_3 \\ C_4\end{pmatrix},\quad\text{for }x_1<0.
\end{equation}
The Fourier transforms of th symmetric and skew-symmetric parts of the weight function, $\jump{\hat{\mathbf{U}}}$ and $\av{\hat{\mathbf{U}}}$, are once 
again given by equations \eqref{Usymft} and \eqref{Uskewft}. However, due to $\mathbf{H}$, and therefore $\mathbf{W}$ being purely real the expressions 
simplify to 
\begin{equation}\label{Usymfthreal}
\jump{\hat{\mathbf{U}}}^+ (\xi)= -\frac{1}{|\xi|}\mathbf{H}\hat{\mathbf{\Sigma}}^-(\xi),
\end{equation}
\begin{equation}\label{Uskewfthreal}
\av{\hat{\mathbf{U}}}(\xi) = -\frac{1}{2|\xi|}\mathbf{W}\hat{\mathbf{\Sigma}}^-(\xi),
\end{equation}
where $\hat{\mathbf{\Sigma}}$ is the $2\times 2$ matrix consisting of two independent tractions. The linearly independent tractions are given by the 
case $C_3=1,C_4=0$ and $C_3=0,C_4=1$ in equation \eqref{parx3sigma}. The results obtained are;
\begin{equation}
\mathbf{\Sigma}^3(x_1)=\frac{(-x_1)^{-\frac{3}{2}}}{\sqrt{2\pi}}\begin{pmatrix}1 \\ 0\end{pmatrix}, \quad
\mathbf{\Sigma}^4(x_1)=\frac{(-x_1)^{-\frac{3}{2}}}{\sqrt{2\pi}}\begin{pmatrix}0 \\ 1\end{pmatrix}.
\end{equation}
The Fourier transforms of these vectors, which are clearly very important in deriving the weight functions, are given by
\begin{equation}
\hat{\mathbf{\Sigma}}^3(\xi)=(i\xi)^{\frac{1}{2}}\begin{pmatrix}-\sqrt{2}\\0\end{pmatrix}, \quad
\hat{\mathbf{\Sigma}}^4(\xi)=(i\xi)^{\frac{1}{2}}\begin{pmatrix}0\\-\sqrt{2}\end{pmatrix}.
\end{equation}

\section{Integral identities}
In this Section the obtained weight function matrices are used together with the Betti identity \eqref{bettipiezo} to formulate the considered crack problem in
terms of singular integral equations. Integral identities relating the applied loading to the resulting crack opening and traction ahead of the tip are derived for transversely
isotropic piezoelectric materials in both the cases where poling direction is parallel to the $x_{2}$ and $x_{3}$ axes.

\subsection{Poling direction parallel to the $x_2-$axis}
Considering the case where both
upper and lower transversely isotropic piezoelectric half-spaces possess poling direction parallel to the $x_2-$axis (perpendicular to the crack plane),
the in-plane fields and piezoelectric effect decouple from the antiplane displacement and traction. Consequently, the Betti formula still has the form
\begin{equation}\label{betti}
\jump{\hat{\mathbf{U}}}^T\bR\hat{\mathbf{t}}^+ - \hat{\mathbf{\Sigma}}^T\bR\jump{\hat{\mathbf{u}}}^- = 
-\jump{\hat{\mathbf{U}}}^T\bR\av{\hat{\mathbf{p}}} - \av{\hat{\mathbf{U}}}^T\bR\jump{\hat{\mathbf{p}}},
\end{equation}
where $\jump{\hat{\mathbf{U}}}$ and $\av{\hat{\mathbf{U}}}$ are given by expressions \eqref{Usymft} and \eqref{Uskewft} together with matrices \eqref{HW3x3},
and the rotational matrix $\mathbf{R}$ becomes
\[
\bR=\begin{pmatrix}-1&0&0\\0&1&0\\0&0&-1\end{pmatrix}.
\]
Multiplying both sides of \eqref{betti} by $\bR^{-1}\jump{\hat{\mathbf{U}}}^{-T}$ yields the following equation
\begin{equation}\label{maineqft}
\hat{\mathbf{t}}^+ - \mathbf{N}\jump{\hat{\mathbf{u}}}^- = -\av{\hat{\mathbf{p}}} - \mathbf{M}\jump{\hat{\mathbf{p}}},
\end{equation}
where $\mathbf{M}$ and $\mathbf{N}$ are given by
\begin{equation}
\mathbf{M}=\bR^{-1}\jump{\hat{\mathbf{U}}}^{-T}\av{\hat{\mathbf{U}}}^T\bR,\qquad
\mathbf{N}=\bR^{-1}\jump{\hat{\mathbf{U}}}^{-T}\hat{\mathbf{\Sigma}}^T\bR.
\end{equation}
Using \eqref{Usymft} and \eqref{Uskewft} full expressions for $\mathbf{M}$ and $\mathbf{N}$ can be found:
\begin{equation}
\mathbf{M}=\frac{1}{2D}\left(\mathbf{M}'+i\mathrm{ sign}(\xi)\mathbf{M}''\right),
\end{equation}\begin{equation}
\mathbf{N}=\frac{|\xi|}{D}\left(\mathbf{N}'+i\mathrm{ sign}(\xi)\mathbf{N}''\right),
\end{equation}
where explicit expressions for $D, \mathbf{M}', \mathbf{M}'', \mathbf{N}'$ and $\mathbf{N}''$ can be found in the Appendix of this paper.

Taking the inverse Fourier transform of equation \eqref{maineqft}, the following equations are found for $x_1<0$ and $x_1>0$ respectively:
\begin{equation}\label{maineqleft}
\mathcal{F}^{-1}_{x_1<0}[\mathbf{N}\jump{\hat{\mathbf{u}}}^-] = \av{\mathbf{p}}(x_1) + \mathcal{F}^{-1}_{x_1<0}[\mathbf{M}\jump{\hat{\mathbf{p}}}],\quad x_1<0,
\end{equation}
\begin{equation}\label{maineqright}
\mathbf{t}^{(+)}(x_1) + \mathcal{F}^{-1}_{x_1>0}[\mathbf{M}\jump{\hat{\mathbf{p}}}] = \mathcal{F}^{-1}_{x_1>0}[\mathbf{N}\jump{\hat{\mathbf{u}}}^-],\quad x_1>0.
\end{equation}
The term involving $\hat{\mathbf{t}}$ cancels for $x_1<0$ as it is only defined along the interface and the $\av{\hat{\mathbf{p}}}$ does not
appear for $x_1>0$ as it is only defined along the crack. In order to derive explicit expressions for equations  \eqref{maineqleft} and \eqref{maineqright}, the inverse Fourier
transform of $i\mathrm{ sign}(\xi)\jump{\hat{\mathbf{u}}}^-$, $|\xi|\jump{\hat{\mathbf{u}}}^-$ and $i \xi\jump{\hat{\mathbf{u}}}^-$ are computed applying the convolutions theorem 
(see Appendix for details).

The singular operator $\mathcal{S}$ and the orthogonal projectors $\mathcal{P}_\pm$ are defined
\begin{equation}
\mathcal{S}\psi = \frac{1}{\pi x_1}\ast\psi(x_1),
\end{equation}
\begin{equation}
\mathcal{P}_\pm\psi=\begin{cases}\psi(x_1),\quad\pm x_1>0,\\0,\quad \text{otherwise}.\end{cases}
\end{equation}
Introducing the singular integral operator $\mathcal{S}^{(s)}=\mathcal{P}_-\mathcal{S}\mathcal{P}_{-}$ 
and the compact operator $\mathcal{S}^{(c)}=\mathcal{P}_+\mathcal{S}\mathcal{P}_{-}$, equations \eqref{maineqleft} and \eqref{maineqright} become
\begin{equation}\label{xless0m}
\mathbf{\mathcal{N}}^{(s)}\frac{\partial \jump{\mathbf{u}}^{(-)}}{\partial x_1} = \av{\mathbf{p}}(x_1) + \mathbf{\mathcal{M}}^{(s)}\jump{\mathbf{p}},\quad x_1<0,
\end{equation}
\begin{equation}\label{xgreater0m}
\mathbf{t}^{(+)}(x_1) + \mathbf{\mathcal{M}}^{(c)}\jump{\mathbf{p}} = \mathbf{\mathcal{N}}^{(c)}\frac{\partial \jump{\mathbf{u}}^{(-)}}{\partial x_1},\quad x_1>0.
\end{equation}
The matrix operators $\mathbf{\mathcal{M}}^{(s)}, \mathbf{\mathcal{M}}^{(c)}, \mathbf{\mathcal{N}}^{(s)}$ and $\mathbf{\mathcal{N}}^{(c)}$ are given by
\begin{equation}
\mathbf{\mathcal{M}}^{(s)}=\frac{1}{2D}\left(\mathbf{M}'+\mathbf{M}''\mathcal{S}^{(s)}\right),
\end{equation}
\begin{equation}
\mathbf{\mathcal{N}}^{(s)}=\frac{1}{D}\left(\mathbf{N}'\mathcal{S}^{(s)}-\mathbf{N}''\right),
\end{equation}
\begin{equation}
\mathbf{\mathcal{M}}^{(c)}=\frac{1}{2D}\mathbf{M}''\mathcal{S}^{(c)},
\end{equation}
\begin{equation}
\mathbf{\mathcal{N}}^{(c)}=\frac{1}{D}\mathbf{N}'\mathcal{S}^{(c)}.
\end{equation}

\subsection{Poling direction parallel to the $x_3-$axis}
For the case where both upper and lower transversely isotropic piezoelectric half-spaces possess poling direction parallel to the $x_3$ axis, the weight functions consist of the
$2\times 2$ matrices \eqref{Usymfthreal} and  \eqref{Uskewfthreal}. The Betti identity \eqref{bettipiezo} then becomes a $2\times 2$ matricial integral equation, where 
the rotational matrix $\mathbf{R}$ is given by
%
%
%
\[
\bR=\begin{pmatrix}-1&0\\0&-1\end{pmatrix}.
\]
Therefore, equation \eqref{bettipiezo} can be simplified further to give
\begin{equation}\label{bettisimp}
\jump{\hat{\mathbf{U}}}^T\hat{\mathbf{t}}^+ - \hat{\mathbf{\Sigma}}^T\jump{\hat{\mathbf{u}}}^- = 
-\jump{\hat{\mathbf{U}}}^T\av{\hat{\mathbf{p}}} - \av{\hat{\mathbf{U}}}^T\jump{\hat{\mathbf{p}}}.
\end{equation}

Multiplying both sides of equation \eqref{bettisimp} by $\jump{\hat{\mathbf{U}}}^{-T}$ gives
\begin{equation}
\hat{\mathbf{t}}^+ - \jump{\hat{\mathbf{U}}}^{-T}\hat{\mathbf{\Sigma}}^T\jump{\hat{\mathbf{u}}}^- = -\av{\hat{\mathbf{p}}} - \jump{\hat{\mathbf{U}}}^{-T}\av{\hat{\mathbf{U}}}^T\jump{\hat{\mathbf{p}}}^-.
\end{equation}
Using \eqref{Usymfthreal} and \eqref{Uskewfthreal} gives
\begin{equation}
\hat{\mathbf{t}}^+ - \mathbf{Z}(\xi)\jump{\hat{\mathbf{u}}}^- = -\av{\hat{\mathbf{p}}} - \mathbf{Y}\jump{\hat{\mathbf{p}}}^-,
\end{equation}
where
\[ \mathbf{Y}= \frac{1}{2}\mathbf{H}^{-1}\mathbf{W},\qquad \mathbf{Z}=-|\xi|\mathbf{H}^{-1}.\]

Taking inverse Fourier transforms and methods seen in \citet{Piccolroaz13} and \citet{Morini2} gives the following singular integral equations
\begin{equation}\label{xless0}
\mathbf{\mathcal{Q}}^{(s)}\frac{\partial\jump{\mathbf{u}}^{(-)}}{\partial x_1}=-\av{\mathbf{p}}(x_1)-\mathbf{Y}\jump{\mathbf{p}}(x_1),\quad\text{for }x_1<0,
\end{equation}
\begin{equation}\label{xgreater0}
\mathbf{t}(x_1)=-\mathbf{\mathcal{Q}}^{(c)}\frac{\partial\jump{\mathbf{u}}^{(-)}}{\partial x_1},\quad\text{for }x_1>0,
\end{equation}
where $\mathbf{\mathcal{Q}}^{(s)}=\mathbf{H}^{-1}\mathbf{\mathcal{S}}^{(s)}$, $\mathbf{\mathcal{Q}}^{(c)}=\mathbf{H}^{-1}\mathbf{\mathcal{S}}^{(c)}$ and
the integral operators $\mathbf{\mathcal{S}}^{(s)}$ and $\mathbf{\mathcal{S}}^{(c)}$ are defined in the same way as those introduced in previous Section.

The integral identities \eqref{xless0m}, \eqref{xgreater0m}, \eqref{xless0} and \eqref{xgreater0} relate the mechanical and electrostatic loading applied on the crack
faces to the corresponding crack opening and tractions ahead of the tip. The crack opening associated with an arbitrary mechanical or electrostatic loading can be 
derived by the inversion of the matricial operators $\mathbf{\mathcal{N}}^{(s)}$ and $\mathbf{\mathcal{Q}}^{(s)}$ in equations \eqref{xless0m} and \eqref{xless0}. Using the obtained
crack opening functions in \eqref{xgreater0m} and \eqref{xgreater0}, explicit expressions for the tractions ahead of the crack tip are yielded. Some simple illustrative examples of this procedure
are reported in next Section. It is important to note that the solution of the obtained systems of singular integral equations
\eqref{xless0m}, \eqref{xgreater0m}, \eqref{xless0} and \eqref{xgreater0}, provides the crack opening displacements, the electric potential on the crack faces, 
the mechanical tractions and electric displacement ahead of the tip associate to general mechanical, electrostatic or electro-mechanical loading without any restriction concerning the
geometry and the symmetry. In particular, the explicit evaluation of the skew-symmetric weight function matrices makes possible to consider the effects of skew-symmetric contributions
to the loading. These effects cannot be accounted by means of other approaches available in literature \citep{Suopiezo, Pak, Pan1, Ma}, which are based on the assumption that the geometry of the 
applied loads is symmetric.


%

\section{Illustrative Examples}
In this Section we consider some examples of loadings for both poling directions and find solutions using the respective singular integral equations derived previously in the paper.
Both mechanical and electrical configurations will be considered. Explicit expressions for crack opening and tractions ahead of the tip corresponding
to both symmetrical and skew-symmetrical mechanical and electrostatic loading configurations are derived. The proposed illustrative cases show that the obtained integral identities 
represent a very useful tool for studying interfacial crack problems in piezoelectric bimaterials. To begin with we consider a symmetric distribution of point loadings when the poling 
direction is parallel to the $x_2$-axis before considering both symmetric an asymmetric loading configurations for the piezoelectric bimaterial poled in the direction of the $x_3$-axis. 
For the decoupled Mode III and IV example with symmetric loading we also present a comparison between the results from our singular integral equations and those obtained using finite 
element methods in COMSOL Multiphysics.

\subsection{Poling direction parallel to the $x_2$-axis under symmetric mechanical loading}
We consider a symmetric system of two perpendicular point loads of varying magnitude on each crack faces acting in the opposite direction to their corresponding load on the opposite 
crack face at a distance $a$ behind the crack. Mathematically these forces are represented as
\begin{equation}
\av{\mathbf{p}}(x_1)=\begin{pmatrix}-F_1\delta(x_1+a) \\ -F_2\delta(x_1+a) \\ 0 \end{pmatrix},\qquad\jump{\mathbf{p}}=\mathbf{0}.
\end{equation}
Under such a loading the singular integral equations used to find $\jump{\mathbf{u}}$ and $\mathbf{t}$ reduce to
\begin{equation}\label{symmsimpcrack}
\mathcal{N}^{(s)}\frac{\partial\jump{\mathbf{u}}^{(-)}}{\partial x_1}=\av{\mathbf{p}}(x_1),\quad\text{for }x_1<0,
\end{equation}
\begin{equation}\label{symmsimpint}
\mathbf{t}^{(+)}(x_1)=\mathcal{N}^{(c)}\frac{\partial\jump{\mathbf{u}}^{(-)}}{\partial x_1},\quad\text{for }x_1>0.
\end{equation}

To simplify the problem we consider the set of bimaterials for which the matrix $\mathbf{H}$ from equation \eqref{HW3x3}$_{(1)}$ has no imaginary part, that is $H_{12}=H_{14}=0$. 
An example of when this may occur is when the upper and lower materials are the same. Under the matrix $\mathbf{N}''=\mathbf{0}$ and therefore the integral equation for $x_1<0$ becomes
\begin{equation}\label{IIx2}
\frac{1}{D}\begin{pmatrix}
N_{11}&0&0\\
0&N_{22}&N_{24}\\
0&N_{24}&N_{44}
\end{pmatrix}\mathcal{S}^{(s)}\frac{\partial\jump{\mathbf{u}}^{(-)}}{\partial x_1}=\begin{pmatrix} -F_1\delta(x_1+a) \\ -F_2\delta(x_1+a) \\ 0 \end{pmatrix}.
\end{equation}
From the system \eqref{IIx2} it is possible to obtain the following three equations for the derivatives of the displacements and electric potential
\begin{equation}\label{m1decouple}
N_{11}\mathcal{S}^{(s)}\frac{\partial\jump{u_1}^{(-)}}{\partial x_1}=-F_1D\delta(x_1+a),
\end{equation}
\begin{equation}\label{m2couplem4}
\mathcal{S}^{(s)}\left[N_{22}\frac{\partial\jump{u_2}^{(-)}}{\partial x_1}+N_{24}\frac{\partial\jump{\phi}}{\partial x_1}\right]=-F_2D\delta(x_1+a),
\end{equation}
\begin{equation}\label{m4couplem2}
\mathcal{S}^{(s)}\left[N_{24}\frac{\partial\jump{u_2}^{(-)}}{\partial x_1}+N_{44}\frac{\partial\jump{\phi}}{\partial x_1}\right]=0.
\end{equation}
It is clear from these equations that the $x_1$ directed part of the solution for the example considered decouples from the component of the solution in the $x_2$ direction and the electrical effects. 
As a result of this we first proceed with solving for $u_1$ before proceeding to find expressions for $u_2$ and $\phi$.

We begin by inverting the integral operator $\mathcal{S}^{(s)}$ in equation \eqref{m1decouple} using the methods seen in \citet{Piccolroaz13} and \citet{Morini2}
\begin{equation}
\frac{\partial\jump{u_1}^{(-)}}{\partial x_1}=\frac{F_1 D}{N_{11}\pi}\int_{-\infty}^0\sqrt{\frac{\eta}{x_1}}\frac{\delta(\eta+a)}{x_1-\eta}\mathrm{d}\eta=\frac{F_1D}{N_{11}\pi}\sqrt{\frac{-a}{x_1}}\frac{1}{x_1+a}.
\end{equation}
Using the expressions for $D$ and $N_{11}$ reported in the Appendix of this paper this expression simplifies to
\begin{equation}\label{m1dispder}
\frac{\partial\jump{u_1}^{(-)}}{\partial x_1}=-\frac{F_1H_{11}}{\pi}\sqrt{-\frac{a}{x_1}}\frac{1}{x_1+a},
\end{equation}
which agrees with those results found in \citet{Piccolroaz13} and \citet{Morini2} when a component of a displacement field decouples from all other components. Note here that the results
differ from those in \citet{Piccolroaz13} due to the anisotropy of the material and they differ from that in \citet{Morini2} due to $u_1$ being decoupled in this paper whereas 
in that paper $u_2$ was separated from the rest of the solution.

Integrating \eqref{m1dispder} gives the following expressions for the displacement jump along the crack
\begin{equation}
\jump{u_1}(x_1)=\frac{2F_1H_{11}}{\pi}\arctanh\sqrt{-\frac{x_1}{a}},\quad\text{for }-a<x_1<0,
\end{equation}\begin{equation}
\jump{u_1}(x_1)=\frac{2F_1H_{11}}{\pi}\arctanh\sqrt{-\frac{a}{x_1}},\quad\text{for }x_1<-a.
\end{equation}

We now proceed to invert the operator in equations \eqref{m2couplem4} and \eqref{m4couplem2} in the same manner. The resulting equations are therefore
\begin{equation}
N_{22}\frac{\partial\jump{u_2}^{(-)}}{\partial x_1}+N_{24}\frac{\partial\jump{\phi}^{(-)}}{\partial x_1}=\frac{F_2D}{\pi}\sqrt{-\frac{a}{x_1}}\frac{1}{x_1+a},
\end{equation}
\begin{equation}
N_{24}\frac{\partial\jump{u_2}^{(-)}}{\partial x_1}+N_{44}\frac{\partial\jump{\phi}^{(-)}}{\partial x_1}=0.
\end{equation}
Solving these equations and simplifying gives the following expressions
\begin{equation}
\frac{\partial\jump{u_2}^{(-)}}{\partial x_1}=-\frac{F_2H_{22}}{\pi}\sqrt{-\frac{a}{x_1}}\frac{1}{x_1+a},
\end{equation}
\begin{equation}
\frac{\partial\jump{\phi}^{(-)}}{\partial x_1}=\frac{F_2H_{24}}{\pi}\sqrt{-\frac{a}{x_1}}\frac{1}{x_1+a}.
\end{equation}
Integrating these expressions gives the following expressions for the jump in displacement and potential over the crack
\begin{equation}
\begin{pmatrix}\jump{u_2}^{(-)}\\\jump{\phi}^{(-)}\end{pmatrix}=\frac{2F_2}{\pi}\arctanh\sqrt{-\frac{x_1}{a}}\begin{pmatrix}H_{22}\\-H_{24}\end{pmatrix},\quad\text{for }-a<x_1<0,
\end{equation}\begin{equation}
\begin{pmatrix}\jump{u_2}^{(-)}\\\jump{\phi}^{(-)}\end{pmatrix}=\frac{2F_2}{\pi}\arctanh\sqrt{-\frac{a}{x_1}}\begin{pmatrix}H_{22}\\-H_{24}\end{pmatrix},\quad\text{for }x_1<-a.
\end{equation}

Using equation \eqref{symmsimpint} the following expressions, for use in finding expressions for the interfacial traction and electric displacement, 
are obtained
\begin{equation}
\begin{pmatrix}\sigma_{21} \\ \sigma_{22} \\ D_2\end{pmatrix}=\frac{1}{D}\begin{pmatrix} N_{11}&0&0\\0&N_{22}&N_{24}\\0&N_{24}&N_{44}\end{pmatrix}
\mathcal{S}^{(c)}\frac{\partial\jump{\mathbf{u}}^{(-)}}{\partial x_1}.
\end{equation}
The decoupled traction component  can now be found:
\begin{equation}
\sigma_{21}^{(+)}(x_1)=\frac{N_{11}}{D\pi}\int_{-\infty}^0\frac{1}{x_1-\eta}\frac{\partial\jump{u_1}^{(-)}}{\partial \eta}\mathrm{d}\eta=\frac{F_1}{\pi}\sqrt{\frac{a}{x_1}}\frac{1}{x_1+a}.
\end{equation}
Once again the obtained result is identical to that obtained for a decoupled field in anisotropic bimaterials \citep{Morini2}, 
with the only difference here arising from the difference in direction of the decoupled field. Using the same method the expressions for the 
coupled portion of the traction and electric displacement field are given as
\begin{equation}
\sigma_{22}(x_1)^{(+)}=\frac{F_2}{\pi}\sqrt{\frac{a}{x_1}}\frac{1}{x_1+a}, \qquad D_2^{(+)}(x_1)=0.
\end{equation}
It is seen that the mechanical part of the solution behaves identically to that in an anisotropic bimaterial and there is no electrical displacement component along the interface for any 
bimaterial with the given conditions.

\subsection{Poling direction parallel to the $x_3$-axis}
\subsubsection{Symmetric mechanical loading}
The loading considered here consists of a point load acting in opposite directions on each of the crack faces at a distance $a$ from the crack tip. Mathematically this system of forces 
is represented using the Dirac delta distribution. The expressions for the symmetric and skew-symmetric parts of the extended loading are given by:
\begin{equation}
\av{\mathbf{p}}(x_1) = \begin{pmatrix} -F\delta(x_1+a) \\ 0\end{pmatrix},\quad \jump{\mathbf{p}}(x_1)=0.
\end{equation}

Inserting these expressions into equation \eqref{xless0} gives the singular integral equation
\begin{equation}
\mathbf{\mathcal{S}}^{(s)}\frac{\partial \jump{\mathbf{u}}^{(-)}}{\partial x_1} = F\begin{pmatrix} H_{33} \\ H_{34} \end{pmatrix} \delta(x_1 + a).
\end{equation}
Inverting the operator $\mathbf{\mathcal{S}}^{(s)}$ in a similar way to that seen previously in the paper gives
\begin{align}
\frac{\partial \jump{\mathbf{u}}^{(-)}}{\partial x_1}  &=-\frac{F}{\pi}\begin{pmatrix} H_{33} \\ H_{34} \end{pmatrix}
\int_{-\infty}^0\sqrt{\frac{\eta}{x_1}}\frac{\delta(\eta +a)}{x_1 - \eta}\mathrm{d}\eta\nonumber\\
 &= 
-\frac{F}{\pi}\sqrt{-\frac{a}{x_1}}\frac{1}{x_1 +a}\begin{pmatrix} H_{33} \\ H_{34} \end{pmatrix}.
\end{align}
Integrating this equation gives the result
\begin{equation}
\begin{pmatrix}\jump{u_3}\\ \jump{\phi}\end{pmatrix} = \frac{2F}{\pi}\arctanh\sqrt{-\frac{x_1}{a}}\begin{pmatrix} H_{33} \\ H_{34} \end{pmatrix},\quad\text{for }-a<x_1<0,
\end{equation}
\begin{equation}
\begin{pmatrix}\jump{u_3}\\ \jump{\phi}\end{pmatrix} = \frac{2F}{\pi}\arctanh\sqrt{-\frac{a}{x_1}}\begin{pmatrix} H_{33} \\ H_{34} \end{pmatrix},\quad\text{for }x_1<-a.
\end{equation}

Making use of equation \eqref{xgreater0} it is possible to obtain the expression for the extended traction vector, $\mathbf{t}$, along the interface:
\begin{align}
\mathbf{t}^{(+)}(x_1)&=-\frac{\mathbf{H}^{-1}}{\pi}\int_{-\infty}^0\frac{1}{x_1-\eta}\frac{\partial\jump{\mathbf{u}}}{\partial\eta}\mathrm{d}\eta,\nonumber\\
&=\frac{F}{\pi}\sqrt{\frac{a}{x_1}}\frac{1}{x_1+a}\mathbf{H}^{-1}\begin{pmatrix}H_{33}\\H_{34}\end{pmatrix},\nonumber\\
&=\frac{F}{\pi}\sqrt{\frac{a}{x_1}}\frac{1}{x_1+a}\begin{pmatrix}1\\0\end{pmatrix}.
\label{symtrac}\end{align}

\vspace{2mm}
\begin{figure}[!htcb]
\centering
\includegraphics[width=13.9cm]{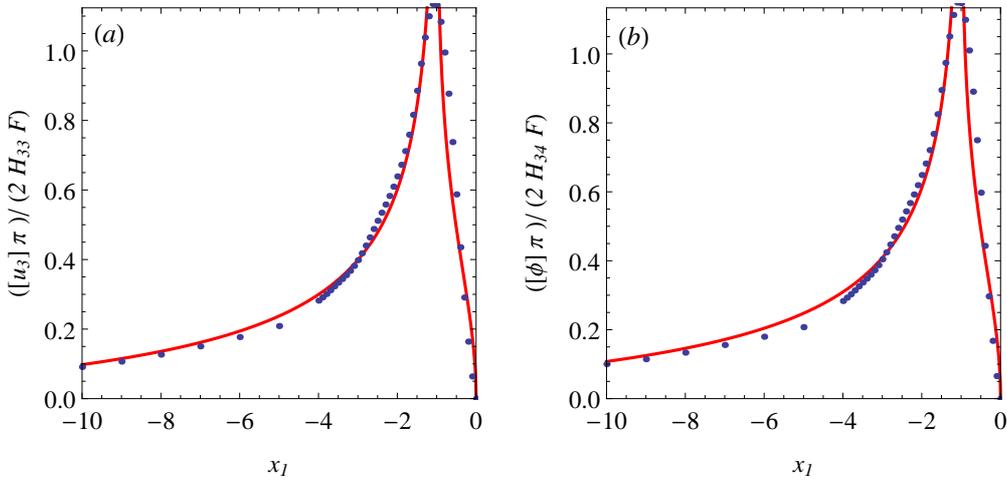}
\caption{\footnotesize (a): Normalized displacement jump associate to symmetric mechanical loading localized in $x_1=a=-1$.
Blue dots are COMSOL multiphysics results;  (b): Normalized electric potential jump associate to symmetric mechanical loading 
localized in $x_1=a=1$. Blue dots are COMSOL multiphysics results.}
\label{figjumpmech}
\end{figure}
\begin{figure}[!htcb]
\centering
\includegraphics[width=13.9cm]{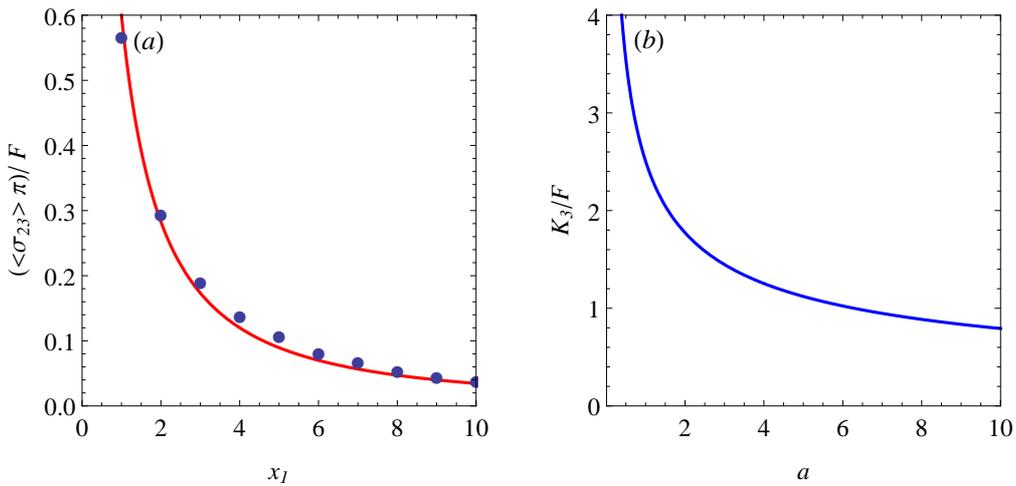}
\caption{ \footnotesize (a): Normalized shear traction associate to symmetric mechanical loading  localized in $x_1=a=-1$.
Blue dots are COMSOL multiphysics results;  (b):  
Variation of the normalized stress intensity factor $K_3$ with the distance $a$ between te point of application of the load and the crack tip.}
\label{figtractK3mech}
\end{figure}
It is now possible to use \eqref{symtrac} to obtain expressions for the stress intensity factor, $K_3$, and the electric intensity factor, $K_4$:
\begin{equation}
\begin{pmatrix}K_3\\K_4\end{pmatrix} = \lim_{x_1\to0}\sqrt{2\pi x_1}\begin{pmatrix}\sigma_{23}\\D_2\end{pmatrix}=F\sqrt{\frac{2}{\pi a}}\begin{pmatrix}1\\0\end{pmatrix}.
\end{equation}

Figures \ref{figjumpmech} and \ref{figtractK3mech}$/(b)$ show a comparison between the derived results and the equivalent results using finite element computations in COMSOL multiphysics. 
The point loads are assumed to be localized at a distance $a=-1$ from the crack tip, and the normalized crack opening, electric potential jump and shear traction
profiles are reported as functions of the spatial coordinate $x_1$. The materials used above and 
below the crack were Barium Titanate and PZT respectively. The material parameters are quoted in Table \ref{matpartable}, with those for Barium
titanate obtained from \citet{Geis} and those for PZT taken from \citet{LiuHsia}. \\
\begin{table}[ht]
\begin{center}
\begin{tabular}{| l | c | c | c |}
\hline
Material & $C_{44}$(GPa) & $e_{15}$(CM$^2$) & $\omega_{11}$(C$^2$/Nm$^2$ \\ \hline
Barium Titanate & 44 & 11.4 & 9.87 x 10$^{-9}$ \\ \hline
PZT & 24.5 & 14.0 & 1.51 x 10$^{-8}$ \\ \hline
\end{tabular}
\caption{Material properties}
\label{matpartable}
\end{center}
\end{table}

Good agreement between the analytical solution and the results provided by finite element analysis is detected in the figures for  normalized crack opening, electric potential jump and shear 
traction field. The variation of the normalized stress intensity factor $K_{3}$ with the distance $a$ bewteen the crack tip and the point where the loads
are applied is reported in Figure$/(b)$.
\subsubsection{Asymmetric mechanical loading}
The second example considered has point loadings at a distance $a$ acting on the upper and lower crack faces. However, for this asymmetric example it is said that they both 
act in the same direction (see for example \citet{Morini2}). Mathematically this is presented as:
\begin{equation}
\av{\mathbf{p}}(x_1)=0,\quad \jump{\mathbf{p}}(x_1) = \begin{pmatrix} -2F\delta(x_1+a) \\ 0\end{pmatrix}.
\label{antimech}
\end{equation}
It is important to note that, as just mentioned, most of the formulations 
previously proposed for fracture mechanics in piezoelectric bimaterials (see for example \citet{Suopiezo, Pak, Pan1, Ma}), do not take into account the presence of skew-symmetric 
components of the load acting on the crack faces. Consequently, in order to solve crack problems involving non-symmetric loading distribution such as the (\ref{antimech}), in Section \ref{WFsec}
we have derived explicit expessions for the skew-symmetric weight functions matrices.

Taking into account the loading distribution (\ref{antimech}), the singular integral equation (\ref{xless0}) becomes
\begin{equation}
\mathbf{\mathcal{S}}^{(s)}\frac{\partial \jump{\mathbf{u}}^{(-)}}{\partial x_1} = -\frac{1}{2}\mathbf{W}\jump{\mathbf{p}}.
\end{equation}
Using the same method as was used for the symmetric loading previously considered it is shown that
\begin{equation}
\begin{pmatrix}\jump{u_3}\\ \jump{\phi}\end{pmatrix} = \frac{2F}{\pi}\arctanh\sqrt{-\frac{x_1}{a}}\begin{pmatrix} \delta_3H_{33} \\ \gamma H_{34} \end{pmatrix},\quad\text{for }-a<x_1<0,
\end{equation}
\begin{equation}
\begin{pmatrix}\jump{u_3}\\ \jump{\phi}\end{pmatrix} = \frac{2F}{\pi}\arctanh\sqrt{-\frac{a}{x_1}}\begin{pmatrix} \delta_3 H_{33} \\ \gamma H_{34} \end{pmatrix},\quad\text{for }x_1<-a.
\end{equation}

Making use of equation \eqref{xgreater0} it is possible to obtain the expression for the extended traction vector, $\mathbf{t}$, along the interface:
\begin{align}
\mathbf{t}^{(+)}(x_1)&=-\frac{\mathbf{H}^{-1}}{\pi}\int_{-\infty}^0\frac{1}{x_1-\eta}\frac{\partial\jump{\mathbf{u}}}{\partial\eta}\mathrm{d}\eta,\nonumber\\
&=\frac{F}{\pi}\sqrt{\frac{a}{x_1}}\frac{1}{x_1+a}\mathbf{H}^{-1}\begin{pmatrix}\delta_3 H_{33}\\\gamma H_{34}\end{pmatrix},\nonumber\\
&=\frac{F}{\pi}\sqrt{\frac{a}{x_1}}\frac{1}{x_1+a}\frac{1}{H_{33}H_{44}-H_{34}^2}\begin{pmatrix}\delta_3H_{33}H_{44}-\gamma H_{34}^2\\(\gamma-\delta_3)H_{33}H_{34}\end{pmatrix}.
\label{asymtrac}\end{align}
It is now possible to use \eqref{asymtrac} to obtain expressions for the stress intensity factor, $K_3$, and the electric intensity factor, $K_4$:
\begin{equation}
\begin{pmatrix}K_3\\K_4\end{pmatrix} = \lim_{x_1\to0}\sqrt{2\pi x_1}\begin{pmatrix}t_3\\D\end{pmatrix}=F\sqrt{\frac{2}{\pi a}}\frac{1}{H_{33}H_{44}-H_{34}^2}\begin{pmatrix}\delta_3H_{33}H_{44}-\gamma H_{34}^2\\(\gamma-\delta_3)H_{33}H_{34}\end{pmatrix}.
\end{equation}

\subsubsection{Symmetric electrical loading}
We consider a symmetric system of electrical point loads on the crack faces at a distance $a$ behind the crack tip. Mathematically the Dirac delta distribution is once again to represent 
the forces:
\begin{equation}
\av{\mathbf{p}}(x_1)=\begin{pmatrix}0\\-G\delta(x_1+a)\end{pmatrix},\quad\jump{\mathbf{p}}=\mathbf{0}.
\end{equation}

Making use of equation \eqref{xless0} and the method previously used for mechanical loading gives
\begin{equation}
\frac{\partial\jump{\mathbf{u}}^{(-)}}{\partial x_1}=-\frac{G}{\pi}\sqrt{-\frac{a}{x_1}}\frac{1}{x_1+a}\begin{pmatrix}H_{34}\\H_{44}\end{pmatrix}.
\end{equation}
When integrated, this gives
\begin{equation}
\begin{pmatrix}\jump{u_3}\\ \jump{\phi}\end{pmatrix} = \frac{2G}{\pi}\arctanh\sqrt{-\frac{x_1}{a}}\begin{pmatrix} H_{34} \\ H_{44} \end{pmatrix},\quad\text{for }-a<x_1<0,
\end{equation}
\begin{equation}
\begin{pmatrix}\jump{u_3}\\ \jump{\phi}\end{pmatrix} = \frac{2G}{\pi}\arctanh\sqrt{-\frac{a}{x_1}}\begin{pmatrix} H_{34} \\ H_{44} \end{pmatrix},\quad\text{for }x_1<-a.
\end{equation}

The resulting expressions for the extended traction vector is therefore
\begin{equation}
\mathbf{t}^{(+)}(x_1)=-\frac{\mathbf{H}^{-1}}{\pi}\int_{-\infty}^0\frac{1}{x_1-\eta}\frac{\partial\jump{\mathbf{u}}^{(-)}}{\partial\eta}\mathrm{d}\eta = \frac{G}{\pi}\sqrt{\frac{a}{x_1}}\frac{1}{x_1+a}\begin{pmatrix}0\\1\end{pmatrix}.
\end{equation}
The stress intensity factors are then given by
\begin{equation}
\begin{pmatrix}K_3\\K_4\end{pmatrix}=\lim_{x_1\to 0}\sqrt{2\pi x_1}\mathbf{t}^{(+)}(x_1) = G\sqrt{\frac{2}{\pi a}}\begin{pmatrix}0\\1\end{pmatrix}.
\end{equation}

\subsubsection{Asymmetric electrical loading}
Here we consider electrical loading acting in the same direction on the top and bottom crack faces at a distance $a$ behind the crack tip. This loading can be written as
\begin{equation}
\av{\mathbf{p}}(x_1)-\mathbf{0},\quad\jump{\mathbf{p}}(x_1)=\begin{pmatrix}0\\-2G\delta(x_1+a)\end{pmatrix}.
\label{antielectro}
\end{equation}
Similarly to the case of asymmetric mechanical load (\ref{antimech}), in order to derive the crack opening and the traction fields ahead of the tip corresponding 
to the asymmetric electrical loading distribution (\ref{antielectro}), the approaches proposed by \citet{Suopiezo, Pak, Pan1, Ma} need to be generalized to the case of non-symmetric
electrical loading applied to the crack faces. As a consequence, the skew-symmetric weight function matrices derived in Section \ref{WFsec} are required. Assuming that the loading function is given by
expression (\ref{antielectro}), the resulting fields obtained using equations \eqref{xless0} and \eqref{xgreater0} are:
\begin{equation}
\frac{\partial\jump{\mathbf{u}}^{(-)}}{\partial x_1}=-\frac{G}{\pi}\sqrt{-\frac{a}{x_1}}\frac{1}{x_1+a}\begin{pmatrix}\gamma H_{34}\\\delta_4 H_{44}\end{pmatrix},
\end{equation}
\begin{equation}
\begin{pmatrix}\jump{u_3}\\ \jump{\phi}\end{pmatrix} = \frac{2G}{\pi}\arctanh\sqrt{-\frac{x_1}{a}}\begin{pmatrix} \gamma H_{34} \\ \delta_4 H_{44} \end{pmatrix},\quad\text{for }-a<x_1<0,
\end{equation}
\begin{equation}
\begin{pmatrix}\jump{u_3}\\ \jump{\phi}\end{pmatrix} = \frac{2G}{\pi}\arctanh\sqrt{-\frac{a}{x_1}}\begin{pmatrix} \gamma H_{34} \\ \delta_4 H_{44} \end{pmatrix},\quad\text{for }x_1<-a.
\end{equation}
\begin{equation}
\mathbf{t}^{(+)}(x_1)=\frac{G}{\pi}\sqrt{\frac{a}{x_1}}\frac{1}{(x_1+a)(H_{33}H_{44}-H_{34}^2)}\begin{pmatrix}(\gamma-\delta_4)H_{34}H_{44}\\ \delta_4 H_{33}H_{44}-\gamma H_{34}^2\end{pmatrix}.
\end{equation}
The stress intensity factor and electric intensity factor obtained are therefore
\begin{equation}
\begin{pmatrix}K_3\\K_4\end{pmatrix}=\lim_{x_1\to 0}\sqrt{2\pi x_1}\mathbf{t}^{(+)}(x_1) = G\sqrt{\frac{2}{\pi a}}\frac{1}{H_{33}H_{44}-H_{34}^2}\begin{pmatrix}(\gamma-\delta_4)H_{34}H_{44}\\ \delta_4 H_{33}H_{44}-\gamma H_{34}^2\end{pmatrix}.
\end{equation}

\section{Conclusions}
A general approach for the derivation of the symmetric and skew-symmetric weight functions for
plane interfacial cracks in anisotropic piezoelectric bimaterials have been developed. The method proposed by \citet{Morini} and \citet{Pryce1} 
for anisotropic elastic bodies, based on the Stroh formalism and Riemann-Hilbert formulation, has been extended for studying crack problems at
the interface between dissimilar piezoelectric media. Applying this approach, explicit weight function matrices are obtained for an
interfacial crack between two transversely isotropic piezoelectric materials, considering both the case where the poling direction of the
two materials is perpendicular and coincident to the crack front. Since many poled ceramics that are commonly used in industrial
applications possess transversely isotropic symmetry, this class of piezoelectric materials has a practical significance,
and the derived weight functions can be used for computing the stress intensity factors corresponding
to any arbitrary non-symmetric mechanical and electrostatic load acting on the crack faces \citep{Piccolroaz09}.

Symmetric and skew-symmetric weight function matrices are used together with Betti's reciprocity theorem to 
derive integral identities relating the applied loading functions to the corresponding crack opening and tractions ahead of the tip.
By means of the proposed method, an explicit singular integral formulation for the crack problem is obtained avoiding the use of Green's
function and the challenging numerical calculations related to such an approach. Integral identities have been derived for interfacial
crack problems between dissimilar transversely isotropic piezoceramics subject to the two-dimensional
state of plane strain and short circuit, having poling direction perpendicular or coincident to the crack front.
An example of the application of the integral identities to crack problems where the loading is given by punctual
forces acting on the faces has been performed. Explicit expressions
for crack opening and tractions ahead of the tip corresponding to
both symmetric and skew-symmetric mechanical and electrical loading have been obtained. The
stress intensity factors associated with the introduced loading configurations are also evaluated. Using the derived skew-symmetric
weight functions matrices, in the illustrative examples the effects of asymmetric loading configurations, which cannot be considered
applying the alternative formulations available in literature, are studied. For the case where the poling direction is perpendicular to the $x_{3}-$axis and symmetric loading is applied at the faces,
the results obtained by the solution of the singular integral equations are compared with those performed by finite
element analysis using COMSOL Multiphysics. Good agreement between the analytical expression for the crack opening obtained
by the singular integral identities and the numerical results is detected.

Thanks to the versatility of the Stroh formalism and the generality of Betti's theorem,
the developed approach can be easily adapted for studying 
several fracture problems in piezoelectric materials with many different properties 
without any assumption concerning the symmetry of the applied loads. 
Furthermore, the derived integral identities
also have their own value from the mathematical point of view, as,
to the authors best knowledge, such identities written in a similar
explicit form for interfacial cracks in anisotropic piezoelectric bimaterials seem
to be unknown in the literature.

\section*{Acknowledgements}
LP, DA and AZ acknowledge support from the FP7 IAPP project `INTERCER2', project reference PIAP-GA-2011-286110-INTERCER2. LM gratefully thanks financial support from the Italian Ministry of Education,
University and Research in the framework of the FIRB project 2010 "Structural mechanics models for renewable energy applications" and from National Group of Mathematical Physics GNFM-INDAM (prot. U2015/000213).

\cleardoublepage

\bibliography{sources}
\bibliographystyle{unsrtnat}
\appendix

\section{Derivation of the Betti formula for piezoelectric materials}
\label{piezoBetti}
In this Appendix the derivation of the generalized Betti identity for piezoelectric materials is reported.  Two sets of stresses, strains, electric fields and electrical displacements
acting on the same physical space are assumed. The two sets of fields are denoted by the superscripts $^{(1)}$ and $^{(2)}$, respectively.
The equations relating these fields are \citep{Hadpiezo}:
\begin{equation}\label{betstart}
\sigma_{ij}^{(1)}\varepsilon_{ij}^{(2)} - D_j^{(1)}E_j^{(2)} = \sigma_{ij}^{(2)}\varepsilon_{ij}^{(1)} - D_j^{(2)}E_j^{(1)}.
\end{equation}
Taking the integral of \eqref{betstart} over a volume, $V$, in combination with equation \eqref{strainfield} yields
\begin{equation}\label{bettivolume}
\int_V\left[(\sigma_{ij}^{(1)}u_i^{(2)})_{,j} + (D_j^{(1)}\phi^{(2)})_{,j}\right]\mathrm{d}V = \int_V\left[(\sigma_{ij}^{(2)}u_i^{(1)})_{,j} + (D_j^{(2)}\phi^{(1)})_{,j}\right]\mathrm{d}V.
\end{equation}
Making use of the Divergence Theorem and then rearranging gives
\begin{equation}\label{BettiS}
\int_{S}\left[\sigma_{ji}^{(1)}n_ju_i^{(2)} + D_j^{(1)}n_j\phi^{(2)} - \sigma_{ji}^{(2)}n_ju_i^{(1)} - D_j^{(2)}n_j\phi^{(1)}\right]\mathrm{d}S = 0, 
\end{equation}
where $S$ is the boundary of the volume $V$. 

Taking $V$ to be a hemisphere in the upper half-plane with flat edge along the $x_2-$plane in equation \eqref{BettiS} leads to the following equation: 
\begin{equation}
\int_{x_2=0^+}\left[\sigma_{2i}^{(1)}u_i^{(2)} + D_2^{(1)}\phi^{(2)} - \sigma_{2i}^{(2)}u_i^{(1)} - D_2^{(2)}\phi^{(1)}\right]\mathrm{d}x_1 = 0, 
\end{equation}
which can written in terms of the extended displacement and traction vectors used in this paper
\begin{equation}
\int_{x_2=0^+}\left[\mathbf{t}^{(1)}\cdot\mathbf{u}^{(2)} - \mathbf{t}^{(2)}\cdot\mathbf{u}^{(1)}\right]\mathrm{d}x_1.
\end{equation}
To obtain equation \eqref{bettiup} $\mathbf{u}^{(1)}$ and $\mathbf{t}^{(1)}$ are taken as the physical fields and then $\mathbf{u}^{(2)}=\mathbf{R}\mathbf{U}$ and 
$\mathbf{t}^{(2)}=\mathbf{R}\mathbf{\Sigma}$. Equation \eqref{bettidown} is derived using a semi-circular domain in the lower half-plane.

\section{Transversely isotropic piezoelectric materials: explicit expressions for matrices $\mathbf{Q}$, $\mathbf{R}$ and $\mathbf{T}$}
In this Appendix, explicit expressions for the stiffness, permittivity and piezoelectric tensors corresponding to transversely isotropic piezoelectric materials
with poling direction parallel to the $x_{2}$ and $x_{3}$ axes are reported. Using these tensors, general forms for the Stroh's eigenvalues matrices, the surface admittance tensor and the bimaterial
matrices $\mathbf{H}$ and  $\mathbf{W}$ are obtained.
\subsection{Poling direction parallel to the $x_2-$axis}
When considering transverse isotropic materials with poling direction parallel to the $x_2$-axis the stiffness tensor, $\mathbf{C}$, simplifies to
\[
\mathbf{C}=\begin{pmatrix}
C_{11}&C_{12}&C_{13}&0&0&0\\
C_{12}&C_{22}&C_{12}&0&0&0\\
C_{13}&C_{12}&C_{11}&0&0&0\\
0&0&0&C_{44}&0&0\\
0&0&0&0&(C_{11}-C_{13})/2&0\\
0&0&0&0&0&C_{44}
\end{pmatrix},
\]
and the permittivity and piezoelectric tensors simplify to
\[
\mathbf{\omega}=\begin{pmatrix}
\omega_{11}&0&0\\
0&\omega_{22}&0\\
0&0&\omega_{11}\end{pmatrix},\qquad
\mathbf{e}=\begin{pmatrix}
0&0&0&0&0&e_{16}\\
e_{21}&e_{22}&e_{21}&0&0&0\\
0&0&0&e_{16}&0&0\end{pmatrix}.
\]
This is the same system as used in \citet{Hwupiezo}. Using these conditions the matrices from equation (\ref{eigenpiezo}) reduce to
\[
\mathbf{Q}=\begin{pmatrix} C_{11}&0&0&0\\
0&C_{44}&0&e_{16}\\0&0&(C_{11}-C_{13})/2&0\\0&e_{16}&0&-\omega_{11}\end{pmatrix},
\quad
\mathbf{R}=\begin{pmatrix} 0&C_{12}&0&e_{21}\\
C_{44}&0&0&0\\0&0&0&0\\e_{16}&0&0&0\end{pmatrix},
\]\[
\mathbf{T}=\begin{pmatrix} C_{44}&0&0&0\\
0&C_{22}&0&e_{22}\\0&0&C_{44}&0\\0&e_{22}&0&-\omega_{22}\end{pmatrix}.
\]
Note that in this case the poling direction is perpendicular to the crack plane reported in Fig. \ref{geometry}.
The antiplane component can clearly be decoupled from the rest of the elasticity components and all effects caused by the electric field. 
This means that the Mode III tractions and displacement have no dependency on the electric field and therefore behave in the same way as they would
in an elastic material with no piezoelectric effect.

Only the in-plane components and electrical effects are considered.
That is: $\mathbf{u}=(u_1,u_2,\phi)^T$ and $\mathbf{t}=(\sigma_{21},\sigma_{22},D_2)^T$. 
The decoupled part of the eigenvalue problem (\ref{eigenpiezo}) now has matrices
\begin{equation}
\mathbf{Q}=\begin{pmatrix} C_{11}&0&0\\
0&C_{44}&e_{16}\\0&e_{16}&-\omega_{11}\end{pmatrix},
\quad
\mathbf{R}=\begin{pmatrix} 0&C_{12}&e_{16}\\
C_{44}&0&0\\e_{21}&0&0\end{pmatrix},
\nonumber\end{equation}\begin{equation}
\mathbf{T}=\begin{pmatrix} C_{44}&0&0\\
0&C_{11}&e_{22}\\0&e_{22}&-\omega_{22}\end{pmatrix}.
\end{equation}

The extended Stroh formalism described in Section 2 of the paper is an effective tool for finding an expression for the surface admittance tensor, $\mathbf{B}$.
However, the eigenvalue problem obtained is not always straightforward to solve. For non-piezoelectric materials an alternative approach is the Lekhnitskii formalism 
introduced by \citet{Lek} which gives a specific normalisation of the eigenvalues obtained from the eigenvalue problem \eqref{eigenpiezo} \citep{Ting}. 
The Lekhnitskii formalism was extended to the piezoelectric setting by \citet{Hwupiezo} where it was used to find the surface admittance tensor for the poling 
direction parallel to the $x_2$-axis

\begin{equation}
\mathbf{B}=\begin{pmatrix}
B_{11}&iB_{12}&iB_{14}\\
-iB_{12}&B_{22}&B_{24}\\
-iB_{14}&B_{24}&B_{44}
\end{pmatrix}.\end{equation}
The expressions for the components of the matrix $\mathbf{B}$ found by \citet{Hwupiezo} are quoted in the Appendix C.

With an expression for $\mathbf{B}$ it is now possible to construct the bimaterial matrices required. The bimaterial matrix $\mathbf{H}$ can be written as
\begin{equation}\label{H3x3ap}
\mathbf{H}=\begin{pmatrix}
H_{11}&iH_{12}&iH_{14}\\
-iH_{12}&H_{22}&H_{24}\\
-iH_{14}&H_{24}&H_{44}
\end{pmatrix},\end{equation}
where
\[H_{\alpha\alpha}=[B_{\alpha\alpha}]_I + [B_{\alpha\alpha}]_{II},\quad\text{for }\alpha=1,2,4,\]
\[H_{1\beta}=[B_{1\beta}]_I - [B_{1\beta}]_{II},\quad\text{for }\beta=2,4,\]
\[H_{24}=[B_{24}]_I + [B_{24}]_{II}.\]
The matrix $\mathbf{W}$ has the same structure as $\mathbf{H}$, that is
\begin{equation}\label{W3x3ap}
\mathbf{W}=\begin{pmatrix}
W_{11}&iW_{12}&iW_{14}\\
-iW_{12}&W_{22}&W_{24}\\
-iW_{14}&W_{24}&W_{44}
\end{pmatrix},\end{equation} 
where
\[W_{\alpha\alpha}=[B_{\alpha\alpha}]_I - [B_{\alpha\alpha}]_{II},\quad\text{for }\alpha=1,2,4,\]
\[W_{1\beta}=[B_{1\beta}]_I + [B_{1\beta}]_{II},\quad\text{for }\beta=2,4,\]
\[W_{24}=[B_{24}]_I - [B_{24}]_{II}.\]

\subsection{Poling direction parallel to the $x_3-$axis}
When considering transverse isotropic materials with poling direction parallel to the $x_3$-axis the stiffness tensor, $C$, simplifies to
\[
C=\begin{pmatrix}
C_{11}&C_{12}&C_{13}&0&0&0\\
C_{12}&C_{11}&C_{13}&0&0&0\\
C_{13}&C_{13}&C_{33}&0&0&0\\
0&0&0&C_{44}&0&0\\
0&0&0&0&C_{44}&0\\
0&0&0&0&0&(C_{11}-C_{12})/2
\end{pmatrix},
\]
and the permittivity and piezoelectric tensors simplify to
\[
\omega=\begin{pmatrix}
\omega_{11}&0&0\\
0&\omega_{11}&0\\
0&0&\omega_{33}\end{pmatrix},\qquad
e=\begin{pmatrix}
0&0&0&0&e_{15}&0\\
0&0&0&e_{15}&0&0\\
e_{31}&e_{31}&e_{33}&0&0&0\end{pmatrix}.
\]
Using these conditions the matrices from equation (\ref{eigenpiezo}) reduce to
\[
\mathbf{Q}=\begin{pmatrix} C_{11}&0&0&0\\
0&(C_{11}-C_{12})/2&0&0\\0&0&C_{44}&e_{15}\\0&0&e_{15}&-\omega_{11}\end{pmatrix},
\quad
\mathbf{R}=\begin{pmatrix} 0&C_{12}&0&0\\
(C_{11}-C_{12})/2&0&0&0\\0&0&0&0\\0&0&0&0\end{pmatrix},
\]\[
\mathbf{T}=\begin{pmatrix} (C_{11}-C_{12})/2&0&0&0\\
0&C_{11}&0&0\\0&0&C_{44}&e_{15}\\0&0&e_{15}&-\omega_{11}\end{pmatrix}.
\]
Note that in this case the poling direction coincides with the front of the crack reported in Fig. \ref{geometry}.
For this particular case it is possible to decouple the Mode I and Mode II components of the displacement and stress fields from the Mode III
fields and piezoelectric effects on the material. This means that the Mode I and II fields will behave similarly to those for purely elastic
materials with no piezoelectric behaviour.

Only the out-of-plane and piezoelectric components are considered. In this case 
$\mathbf{Q}$, $\mathbf{R}$ and $\mathbf{T}$ are reduced to 2x2 matrices:
\begin{equation}
\mathbf{Q}=\mathbf{T}=\begin{pmatrix}
C_{44}&e_{15}\\ e_{15} & -\omega_{11}\end{pmatrix},\qquad
\mathbf{R}=\mathbf{0}.
\end{equation}

As a consequence, for this case the surface admittance tensor, $\mathbf{B}$ assumes the form
\begin{equation}
\mathbf{B}=\begin{pmatrix} B_{33} & B_{34} \\ B_{34} & B_{44} \end{pmatrix}.
\end{equation}
Explicit expressions for the components of $\mathbf{B}$ are given in the Appendix C.

The bimaterial matrices $\mathbf{H}$ and $\mathbf{W}$ can be computed and have the form
\begin{equation}
\mathbf{H}=\begin{pmatrix}H_{33} & H_{34} \\ H_{34} & H_{44} \end{pmatrix},\qquad
\mathbf{W}=\begin{pmatrix} \delta_3 H_{33} & \gamma H_{34} \\ \gamma H_{34} & \delta_4 H_{44} \end{pmatrix}.
\end{equation}
The components of these matrices are given by
\[H_{\alpha\beta}=[B_{\alpha\beta}]_I + [B_{\alpha\beta}]_{II},\quad\text{for }\alpha,\beta=3,4,\]
\[ \delta_\alpha = \frac{[B_{\alpha\alpha}]_I - [B_{\alpha\alpha}]_{II}}{[B_{\alpha\alpha}]_I + [B_{\alpha\alpha}]_{II}},\quad\text{for }\alpha=3,4,\]
\[ \gamma = \frac{[B_{34}]_I - [B_{34}]_{II}}{H_{34}}.\]

\section{Explicit expressions for $\mathbf{B}=i\mathbf{AL}^{-1}$}
In this section we give the expressions for the surface admittance tensor $\mathbf{B}$ corresponding to the two-dimensional state of plane
strain and short circuit ($\varepsilon_{3}=0, E_{3}=0$), as seen in \citet{Hwupiezo}. 
We only give the expressions for the decoupled part of the tensor which contains the piezoelectric behaviour of the material.
\subsection{Poling direction parallel to the $x_2$-axis}
As stated in Section 2, the general form of the matrix $\mathbf{B}=i\mathbf{AL}^{-1}$ for transverse isotropic materials with poling direction parallel to the $x_2$-axis is
\begin{equation}
\mathbf{B}=\begin{pmatrix}
B_{11}&iB_{12}&iB_{14}\\
-iB_{12}&B_{22}&B_{24}\\
-iB_{14}&B_{24}&B_{44}
\end{pmatrix}.\end{equation}
Here, explicit expressions for the components of this matrix (found in \citet{Hwupiezo}) are given.

The following components of the compliance tensor: $\mathbf{S}$, piezoelectric strain/voltage tensor: $\mathbf{g}$, and dielectric non-permittivity: $\mathbf{\beta}$, are introduced:
\begin{align}
\hat{S'}_{11}&=\frac{C_{22}}{C_{11}C_{22}-C_{12}^2} - \frac{(e_{21}C_{11}-e_{22}C_{12})^2}{C^*[C_{11}C_{22}-C_{12}^2]},\nonumber\\
\hat{S'}_{12}&=-\frac{e_{21}e_{22}[C_{11}^2-C_{12}^2]}{C^*[C_{11}C_{22}-C_{12}^2]} + \frac{\omega_{22}C_{12}}{C^*},\nonumber\\
\hat{S'}_{22}&=\frac{e_{21}^2[C_{11}^2-C_{12}^2]}{C^*[C_{11}C_{22}-C_{12}^2]} + \frac{\omega_{22}C_{11}}{C^*},\nonumber\\
\hat{S'}_{66}&=\frac{\omega_{11}}{e_{16}^2+\omega_{11}C_{44}},\nonumber\\
\hat{g'}_{21}&=\frac{e_{21}C_{11} - e_{22}C_{12}}{C^*},\nonumber\\
\hat{g'}_{22}&=\frac{e_{22}C_{11} - e_{21}C_{12}}{C^*},\nonumber\\
\hat{g'}_{16}&=\frac{e_{16}}{e_{16}^2 + \omega_{11}C_{44}},\nonumber\\
\hat{\beta'}_{11}&=\frac{C_{44}}{e_{16}^2 + \omega_{11}C_{44}},\nonumber\\
\hat{\beta'}_{22}&=\frac{C_{11}C_{22}-C_{12}^2}{C^*},\nonumber
\end{align}
where
\[
C^*=(e_{21}^2 + e_{22}^2)C_{11} - 2e_{21}e_{22}C_{12} + \omega_{22}[C_{11}C_{22}-C_{12}^2].
\]

Through using the Lekhnitskii formalism extended to piezoelectric materials the eigenvalues, $p$, are found through the equation 
\begin{equation}\label{tix2eigen}
l_4\rho_2-m_3^2=0,
\end{equation}
where $l_4,\rho_2$ and $m_3$ are functions of $p$ and are given by
\begin{equation}
l_4=\hat{S'}_{11}p^4 + (2\hat{S'}_{12}+\hat{S'}_{66})p^2 + \hat{S'}_{22},\quad
m_3=-(\hat{g'}_{21}+\hat{g'}_{16})p^2 - \hat{g'}_{22},\quad \rho_2=-(\hat{\beta'}_{11}p^2 + \hat{\beta'}_{22}).
\end{equation}
This sextic equation must be solved numerically but is easily shown to have roots of the form
\begin{equation}
p_2=\alpha_2+i\beta_2,\quad p_3=-\alpha_2+i\beta_2,\quad p_4=i\beta_4.
\end{equation}

With the eigenvalues known, \citet{Hwupiezo} proceeded to find explicit expressions for the components of $\mathbf{B}$. It was shown that
\begin{align}
B_{11}&= 2\hat{S'}_{11}\mathrm{ Im}\{p_2^2\bar{\eta_2}+(p_4^2\eta_2-p_2^2\eta_4)\}/\lambda,\nonumber\\
B_{22}&= 2\mathrm{ Im}\{[\gamma_2\bar{p_2}p_4\eta_4 + (\gamma_2p_4^2-\gamma_4p_2^2)\bar{\eta_2}]/p_2p_4\}/\lambda,\nonumber\\
B_{44}&= -2\hat{\beta'}_{11}\mathrm{ Im}\{p_2\bar{p_2}\eta_2 + p_2p_4(\eta_2-\eta_4)\}/\lambda,\nonumber\\
B_{24}&= 2\hat{\beta'}_{11}\mathrm{ Im}\{p_2\bar{p_2}\eta_2\eta_4 + p_2p_4\bar{\eta_2}(\eta_2-\eta_4)\}/\lambda,\nonumber\\
B_{12}&= \hat{S'}_{12} + 2\mathrm{ Re}\{[\gamma_2 p_4\eta_2+(\gamma_4p_2\eta_2-\gamma_2p_4\eta_4)]/p_2p_4\}/\lambda,\nonumber\\
B_{14}&= -\hat{g'}_{16} + 2\hat{\beta'}_{11}\mathrm{ Re}\{p_2\eta_2\bar{\eta_2} - \eta_2\eta_4(p_2-p_4)\}/\lambda,\nonumber
\end{align}
where
\[
\lambda=2\mathrm{ Re}\{\bar{p_2}\eta_2 + (p_4\eta_2-p_2\eta_4)\},
\]\[
\gamma_k = \hat{S'}_{22} + \hat{g'}_{22}\eta_k,\quad \eta_k=\frac{l_4(p_k)}{m_3(p_k)},\quad\text{for k=2,4}.
\]

\subsection{Poling direction parallel to the $x_3$-axis}
The decoupled part of the surface admittance tensor, $\mathbf{B}=i\mathbf{AL}^{-1}$, containing Mode III fields and the piezoelectric effect for transverse isotropic materials with poling direction parallel to the $x_3$-axis is:
\begin{equation}
\mathbf{B}=\begin{pmatrix} 
B_{33} & B_{34}\\
B_{34} & B_{44}
\end{pmatrix}.
\end{equation}
Explicit expressions for the components of this matrix are given by:
\[ B_{33} = \frac{\omega_{11}}{e_{15}^2 + \omega_{11}C_{44}},\]
\[ B_{44} = \frac{-C_{44}}{e_{15}^2 + \omega_{11}C_{44}},\]
\[ B_{34} = \frac{e_{15}}{e_{15}^2 + \omega_{11}C_{44}}.\]

\section{Inverse Fourier transforms}
In order to derive explicit expressions for equations \eqref{maineqleft} and \eqref{maineqright},
we need to compute the inverse Fourier trasform of terms of the form $i\mathrm{ sign}(\xi)\hat{f}(\xi)$, $|\xi|\hat{f}(\xi)$ and $i \xi\hat{f}(\xi)$. 
Using the Fourier convolution theorem and the knowledge that the inverse Fourier transform of $\mathrm{sign}(\xi)$ is given by $-i/(\pi x_1)$ gives:
\begin{align}
\mathcal{F}^{-1}[i\mathrm{ sign}(\xi)\hat{f}(\xi)]&=i\mathcal{F}^{-1}[\mathrm{ sign}(\xi)]\ast\mathcal{F}^{-1}[\hat{f}(\xi)], \nonumber\\
&=i\Big( \frac{-i}{\pi x_1}\Big) \ast f(x_1),\nonumber\\
&=\frac{1}{\pi}\int^\infty_{-\infty}\frac{f(\eta)}{x_1-\eta}\mathrm{d}\eta.
\end{align}
The inverse Fourier transform of $|\xi|\hat{f}(\xi)$ is found:
\begin{align}
\mathcal{F}^{-1}[|\xi|\hat{f}(\xi)]&=\mathcal{F}^{-1}[\mathrm{sign}(\xi)]\ast\mathcal{F}^{-1}[\xi\hat{f}(\xi)],\nonumber\\
&=\Big( -\frac{i}{\pi x_1}\Big)\ast i\frac{\partial f}{\partial x_1},\nonumber\\
&=\Big( \frac{1}{\pi x_1}\Big)\ast \frac{\partial f}{\partial x_1}.
\end{align}
Finally, the inverse Fourier transform of $i\mathrm{sign}(\xi)|\xi|\hat{f}(\xi)$ is given by
\begin{align}
\mathcal{F}^{-1}[i\mathrm{sign}(\xi)|\xi|\hat{f}(\xi)]&=\mathcal{F}^{-1}[i\xi\hat{f}(\xi)],\nonumber\\
&=-\frac{\partial f}{\partial x_1}.
\end{align}

\section{Expressions for matrices $\mathbf{M}$ and $\mathbf{N}$}
In this Appendix explicit expressions for the matrices $\mathbf{M}$ and $\mathbf{N}$ are quoted. They have the form
\begin{equation}
\mathbf{M}=\frac{1}{2D}\left(\mathbf{M}'+i\mathrm{ sign}(\xi)\mathbf{M}''\right),
\end{equation}\begin{equation}
\mathbf{N}=\frac{|\xi|}{D}\left(\mathbf{N}'+i\mathrm{ sign}(\xi)\mathbf{N}''\right),
\end{equation}
where
\begin{equation}
D=H_{14}^2H_{22} + H_{12}^2H_{44} + H_{24}^2H_{11} - H_{11}H_{22}H_{44} - 2H_{14}H_{12}H_{24}.
\end{equation}
The matrices $\mathbf{M}',\mathbf{M}'', \mathbf{N}'$ and $\mathbf{N}''$ have the form
\begin{equation}
\mathbf{M}'=\begin{pmatrix}
M_{11}&0&0\\
0&M_{22}&M_{24}\\
0&M_{42}&M_{44}\end{pmatrix},
\qquad
\mathbf{M}''=\begin{pmatrix}
0&M_{12}&M_{14}\\
M_{21}&0&0\\
M_{41}&0&0\end{pmatrix},
\end{equation}
\begin{equation}
\mathbf{N}'=\begin{pmatrix}
N_{11}&0&0\\
0&N_{22}&N_{24}\\
0&N_{24}&N_{44}\end{pmatrix},
\qquad
\mathbf{N}''=\begin{pmatrix}
0&N_{12}&N_{14}\\
-N_{12}&0&0\\
-N_{14}&0&0\end{pmatrix},
\end{equation}
where
\begin{align}
M_{11}&=W_{11}(H_{24}^2-H_{22}H_{44}) + W_{12}(H_{12}H_{44}-H_{14}H_{24}) - W_{14}(H_{12}H_{24}-H_{14}H_{22}),\\
M_{22}&=W_{12}(H_{12}H_{44}-H_{14}H_{24}) - W_{22}(H_{11}H_{44}-H_{14}^2) - W_{24}(H_{14}H_{12}-H_{11}H_{24}),\\
M_{44}&=W_{14}(H_{14}H_{22}-H_{12}H_{24}) - W_{24}(H_{14}H_{12}-H_{11}H_{24}) - W_{44}(H_{11}H_{22}-H_{12}^2),\\
M_{24}&=W_{14}(H_{14}H_{24}-H_{12}H_{44}) + W_{24}(H_{11}H_{44}-H_{14}^2) + W_{44}(H_{14}H_{12}-H_{11}H_{24}),\\
M_{42}&=W_{12}(H_{12}H_{24}-H_{14}H_{22}) + W_{22}(H_{14}H_{12}-H_{11}H_{24}) + W_{24}(H_{11}H_{22}-H_{12}^2),
\end{align}
\begin{align}
M_{12}&=W_{12}(H_{22}H_{44}-H_{24}^2) - W_{22}(H_{12}H_{44}-H_{14}H_{24}) + W_{24}(H_{12}H_{24}-H_{14}H_{22}),\\
M_{14}&=W_{14}(H_{24}^2-H_{22}H_{44}) + W_{24}(H_{12}H_{44}-H_{14}H_{24}) - W_{44}(H_{12}H_{24}-H_{14}H_{22}),\\
M_{21}&=W_{11}(H_{12}H_{44}-H_{14}H_{24}) - W_{12}(H_{11}H_{44}-H_{14}^2) - W_{14}(H_{14}H_{12}-H_{11}H_{24}),\\
M_{41}&=W_{11}(H_{12}H_{24}-H_{14}H_{22}) + W_{12}(H_{14}H_{12}-H_{11}H_{24}) + W_{14}(H_{11}H_{22}-H_{12}^2),
\end{align}
\begin{align}
N_{11}&= H_{22}H_{44}-H_{24}^2,\\
N_{22}&= H_{11}H_{44}-H_{14}^2,\\
N_{44}&= H_{11}H_{22}-H_{12}^2,\\
N_{24}&= H_{11}H_{24}-H_{14}H_{12},
\end{align}
\begin{align}
N_{12}&= H_{12}H_{44}-H_{14}H_{24},\\
N_{14}&= H_{12}H_{24}-H_{14}H_{22}.
\end{align}

\section{Expressions for matrices $\mathbf{Y}$ and $\mathbf{Z}(\xi)$}
Explicit expressions for the matrices $\mathbf{Y}$ and $\mathbf{Z}(\xi)$ are given by:
\begin{equation}
\mathbf{Y}=\frac{1}{2}\mathbf{H}^{-1}\mathbf{W}=\frac{1}{2(H_{33}H_{44}-H_{34}^2)}\begin{pmatrix}
\delta_3H_{33}H_{44}-\gamma H_{34}^2 & H_{44}H_{34}(\gamma - \delta_4)\\
H_{33}H_{34}(\gamma - \delta_3) & \delta_4H_{33}H_{44}-\gamma H_{34}^2
\end{pmatrix},
\end{equation}
\begin{equation}
\mathbf{Z}(\xi) = -\frac{|\xi|}{H_{33}H_{44}-H_{34}^2} \begin{pmatrix} H_{44} & - H_{34} \\ -H_{34} & H_{33} \end{pmatrix}.
\end{equation}

\end{document}